\RequirePackage{fix-cm}
\documentclass[twocolumn,epjc3]{svjour3}  
\smartqed  
\RequirePackage{graphicx}
\RequirePackage[colorlinks,citecolor=blue,urlcolor=blue,linkcolor=blue]{hyperref} 
\usepackage{color}
\usepackage{multirow}
\usepackage{amssymb}     
\usepackage{marvosym}    
\usepackage{upgreek}     
\newcommand{\ctsper}      {cts/(keV$\cdot$kg$\cdot$yr)}

\newcommand{\mum}         {{$\upmu$m}}

\newcommand{\qbb}         {{$Q_{\beta\beta}$}}

\newcommand{\onbb}        {{$0\nu\beta\beta$}}

\newcommand{\fgesix}      {\mbox{$f_{76}$}}

\newcommand{\factvol}     {\mbox{$f_{av}$}}

\newcommand{\gerda}       {\textsc{Gerda}}

\newcommand{\GERDA}       {\mbox{\textsc{Gerda}}}  
\newcommand{\majorana}    {\textsc{Majorana}}

\newcommand{\hdm}         {\textsc{HdM}}

\newcommand{\geant}       {\textsc{Geant4}}

\newcommand{\mage}        {\textsc{MaGe}}
\newcommand{\gelatio}     {\textsc{Gelatio}}

\newcommand{\gesix}       {{$^{76}$Ge}}

\newcommand{\geenr}       {{$^{\rm enr}$Ge}}          
\newcommand{\genat}       {{$^{\rm nat}$Ge}}
\newcommand{\gedep}       {{$^{\rm dep}$Ge}}

\journalname{Eur. Phys. J. C}

\begin{document}

\title{Production, characterization and operation of\\$^{76}$Ge enriched BEGe
  detectors in {\mbox{{\textsc GERDA}}}}

\titlerunning{enriched BEGe detectors for {\mbox{{\textsc Gerda}}}} 
\author{
M.~Agostini\thanksref{TUM} \and
M.~Allardt\thanksref{DD} \and
E.~Andreotti\thanksref{GEEL,TU} \and
A.M.~Bakalyarov\thanksref{KU} \and
M.~Balata\thanksref{ALNGS} \and
I.~Barabanov\thanksref{INR} \and
N.~Barros\thanksref{DD} \and
L.~Baudis\thanksref{UZH} \and
C.~Bauer\thanksref{HD} \and
N.~Becerici-Schmidt\thanksref{MPIP} \and
E.~Bellotti\thanksref{MIBF,MIBINFN} \and
S.~Belogurov\thanksref{ITEP,INR} \and
S.T.~Belyaev\thanksref{KU} \and
G.~Benato\thanksref{UZH} \and
A.~Bettini\thanksref{PDUNI,PDINFN} \and
L.~Bezrukov\thanksref{INR} \and
T.~Bode\thanksref{TUM} \and
D.~Borowicz\thanksref{CR,JINR} \and
V.~Brudanin\thanksref{JINR} \and
R.~Brugnera\thanksref{PDUNI,PDINFN} \and
D.~Budj{\'a}{\v{s}}\thanksref{TUM} \and
A.~Caldwell\thanksref{MPIP} \and
C.~Cattadori\thanksref{MIBINFN} \and
A.~Chernogorov\thanksref{ITEP} \and
V.~D'Andrea\thanksref{ALNGS,alsoGSSI} \and
E.V.~Demidova\thanksref{ITEP} \and
A.~Domula\thanksref{DD} \and
V.~Egorov\thanksref{JINR} \and
R.~Falkenstein\thanksref{TU} \and
K.~Freund\thanksref{TU} \and
N.~Frodyma\thanksref{CR} \and
A.~Gangapshev\thanksref{INR,HD} \and
A.~Garfagnini\thanksref{PDUNI,PDINFN} \and
C.~Gotti\thanksref{MIBINFN,alsoFI} \and 
P.~Grabmayr\thanksref{TU} \and
V.~Gurentsov\thanksref{INR} \and
K.~Gusev\thanksref{KU,JINR,TUM} \and
W.~Hampel\thanksref{HD} \and
A.~Hegai\thanksref{TU} \and
M.~Heisel\thanksref{HD} \and
S.~Hemmer\thanksref{PDUNI,PDINFN} \and
G.~Heusser\thanksref{HD} \and
W.~Hofmann\thanksref{HD} \and
M.~Hult\thanksref{GEEL} \and
L.V.~Inzhechik\thanksref{INR,alsoMIPT} \and
L.~Ioannucci\thanksref{ALNGS} \and
J.~Janicsk{\'o} Cs{\'a}thy\thanksref{TUM} \and
J.~Jochum\thanksref{TU} \and
M.~Junker\thanksref{ALNGS} \and
V.~Kazalov\thanksref{INR} \and
T.~Kihm\thanksref{HD} \and
I.V.~Kirpichnikov\thanksref{ITEP} \and
A.~Kirsch\thanksref{HD} \and
A.~Klimenko\thanksref{HD,JINR,alsoIUN} \and
K.T.~Kn{\"o}pfle\thanksref{HD} \and
O.~Kochetov\thanksref{JINR} \and
V.N.~Kornoukhov\thanksref{ITEP,INR} \and
V.V.~Kuzminov\thanksref{INR} \and
M.~Laubenstein\thanksref{ALNGS} \and
A.~Lazzaro\thanksref{TUM} \and
V.I.~Lebedev\thanksref{KU} \and
B.~Lehnert\thanksref{DD} \and
H.Y.~Liao\thanksref{MPIP} \and
M.~Lindner\thanksref{HD} \and
I.~Lippi\thanksref{PDINFN} \and
A.~Lubashevskiy\thanksref{HD,JINR} \and
B.~Lubsandorzhiev\thanksref{INR} \and
G.~Lutter\thanksref{GEEL} \and
C.~Macolino\thanksref{ALNGS,alsoGSSI} \and
B.~Majorovits\thanksref{MPIP} \and
W.~Maneschg\thanksref{HD} \and
M.~Misiaszek\thanksref{CR} \and
I.~Nemchenok\thanksref{JINR} \and
S.~Nisi\thanksref{ALNGS} \and
C.~O'Shaughnessy\thanksref{MPIP,nowUNC} \and 
D.~Palioselitis\thanksref{MPIP} \and 
L.~Pandola\thanksref{ALNGS,alsoLNS} \and
K.~Pelczar\thanksref{CR} \and
G.~Pessina\thanksref{MIBF,MIBINFN} \and
A.~Pullia\thanksref{MILUINFN} \and
S.~Riboldi\thanksref{MILUINFN} \and
N.~Rumyantseva\thanksref{JINR} \and
C.~Sada\thanksref{PDUNI,PDINFN} \and
M.~Salathe\thanksref{HD} \and
C.~Schmitt\thanksref{TU} \and
J.~Schreiner\thanksref{HD} \and
O.~Schulz\thanksref{MPIP} \and
B.~Schwingenheuer\thanksref{HD} \and
S.~Sch{\"o}nert\thanksref{TUM} \and
E.~Shevchik\thanksref{JINR} \and
M.~Shirchenko\thanksref{KU,JINR} \and
H.~Simgen\thanksref{HD} \and
A.~Smolnikov\thanksref{HD} \and
L.~Stanco\thanksref{PDINFN} \and
H.~Strecker\thanksref{HD} \and
C.A.~Ur\thanksref{PDINFN} \and
L.~Vanhoefer\thanksref{MPIP} \and
A.A.~Vasenko\thanksref{ITEP} \and
K.~von Sturm\thanksref{PDUNI,PDINFN} \and
V.~Wagner\thanksref{HD} \and
M.~Walter\thanksref{UZH} \and
A.~Wegmann\thanksref{HD} \and
T.~Wester\thanksref{DD} \and
H.~Wilsenach\thanksref{DD} \and
M.~Wojcik\thanksref{CR} \and
E.~Yanovich\thanksref{INR} \and
P.~Zavarise\thanksref{ALNGS} \and
I.~Zhitnikov\thanksref{JINR} \and
S.V.~Zhukov\thanksref{KU} \and
D.~Zinatulina\thanksref{JINR} \and
K.~Zuber\thanksref{DD} \and
G.~Zuzel\thanksref{CR} 
}

\authorrunning{the \textsc{Gerda} collaboration}

\thankstext{alsoGSSI}{\emph{also at:}INFN Gran Sasso Science Institute, L'Aquila, Italy}
\thankstext{alsoFI}{\emph{also at:} Universit{\`a} di Firenze, Firenze, Italy}
\thankstext{alsoMIPT}{\emph{also at:} Moscow Inst. of Physics and Technology,
  Russia} 
\thankstext{alsoIUN}{\emph{also at:} Int. Univ. for Nature, Society and
    Man ``Dubna'', Dubna, Russia} 
\thankstext{nowUNC}{\emph{Present Address:} U. North Carolina, Chapel Hill, USA}
\thankstext{alsoLNS}{\emph{Present Address:}  Laboratori Nazionali del Sud,
  INFN, Catania, Italy}
\thankstext{corrauthor}{\emph{Correspondence},
                                email: gerda-eb@mpi-hd.mpg.de}
\institute{
INFN Laboratori Nazionali del Gran Sasso, LNGS, Assergi, Italy\label{ALNGS} \and
Institute of Physics, Jagiellonian University, Cracow, Poland\label{CR} \and
Institut f{\"u}r Kern- und Teilchenphysik, Technische Universit{\"a}t Dresden,
      Dresden, Germany\label{DD} \and
Joint Institute for Nuclear Research, Dubna, Russia\label{JINR} \and
Institute for Reference Materials and Measurements, Geel,
     Belgium\label{GEEL} \and
Max Planck Institut f{\"u}r Kernphysik, Heidelberg, Germany\label{HD} \and
Dipartimento di Fisica, Universit{\`a} Milano Bicocca,
     Milano, Italy\label{MIBF} \and
INFN Milano Bicocca, Milano, Italy\label{MIBINFN} \and
Dipartimento di Fisica, Universit{\`a} degli Studi di Milano e INFN Milano,
    Milano, Italy\label{MILUINFN} \and
Institute for Nuclear Research of the Russian Academy of Sciences,
    Moscow, Russia\label{INR} \and
Institute for Theoretical and Experimental Physics,
    Moscow, Russia\label{ITEP} \and
National Research Centre ``Kurchatov Institute'', Moscow, Russia\label{KU} \and
Max-Planck-Institut f{\"ur} Physik, M{\"u}nchen, Germany\label{MPIP} \and
Physik Department and Excellence Cluster Universe,
    Technische  Universit{\"a}t M{\"u}nchen, M{\"u}nchen, Germany\label{TUM}
                     \and
Dipartimento di Fisica e Astronomia dell{`}Universit{\`a} di Padova,
    Padova, Italy\label{PDUNI} \and
INFN  Padova, Padova, Italy\label{PDINFN} \and
Physikalisches Institut, Eberhard Karls Universit{\"a}t T{\"u}bingen,
    T{\"u}bingen, Germany\label{TU} \and
Physik Institut der Universit{\"a}t Z{\"u}rich, Z{\"u}rich,
    Switzerland\label{UZH}
}

\date{Received: date / Accepted: date}

\maketitle

\begin{abstract}
The GERmanium Detector Array (\gerda) at the Gran Sasso Underground Laboratory
(LNGS) searches for the neutrinoless double beta decay ($0\nu\beta\beta$) of
$^{76}$Ge. Germanium detectors made of material with an enriched $^{76}$Ge
fraction act simultaneously as sources and detectors for this decay.

During Phase~I of the experiment mainly refurbished semi-coaxial Ge detectors
from former experiments were used. For the upcoming Phase~II, 30 new $^{76}$Ge
enriched detectors of broad energy germanium (BEGe)-type were produced. A
subgroup of these detectors has already been deployed in \gerda\ during
Phase~I.

The present paper reviews the complete production chain of these BEGe
detectors including isotopic enrichment, purification, crystal growth and
diode production. The efforts in optimizing the mass yield and in minimizing
the exposure of the $^{76}$Ge enriched germanium to cosmic radiation during
processing are described. Furthermore, characterization measurements in vacuum
cryostats of the first subgroup of seven BEGe detectors and their long-term
behavior in liquid argon are discussed. The detector performance fulfills the
requirements needed for the physics goals of \gerda\ Phase~II.
      
\end{abstract}

\section{Introduction}
 \label{sec:intro}
 
The GERmanium Detector Array
(\gerda)~\cite{bib:GER13,bib:GER13-0nbb,bib:GER13-bckg,bib:GER13-psd} is an
experiment at the Laboratori Nazionali del Gran Sasso (LNGS) of INFN searching
for the neutrinoless double beta (\onbb) decay of $^{76}$Ge. It uses
high-purity germanium (HPGe) detectors that are enriched in $^{76}$Ge to
(86-88)\% as sources and as detection media. The detectors are mounted in
low-mass holders and are embedded in liquid argon (LAr), which serves as a
cryogenic coolant and absorber against external radiation. A tank filled with
ultrapure water provides a 3\,m thick water buffer around the cryostat and
serves as an additional absorber and as a Cherenkov muon veto.

The experimental signature of \onbb\ decay is a peak in the spectrum of the
summed energies of the two electrons released in the nuclear process. The peak
should arise at the \qbb\ value which in the case of $^{76}$Ge is at
(2039.061$\pm$0.007)\,keV~\cite{bib:qvalue}.  The expected number of signal
events $\lambda_S$ is given by:
\begin{equation}
\lambda_S = \frac{\ln\,2 \cdot t}{T_{1/2}^{0\nu}} \cdot
       \frac{N_A \cdot M} {m_{enr}} \cdot
       \fgesix \cdot \factvol \cdot \varepsilon_{fep}\cdot \varepsilon_{psd}
\label{eq:sensitivity}
\end{equation}
where $T_{1/2}^{0\nu}$ is the half life of the \onbb\ decay, $t$ the live
time of the measurement, $N_A$ the Avogadro constant, $m_{enr}$ the molar
mass of the enriched material and $M$ the total detector mass. The parameters
$\factvol$ and $\fgesix$ correspond to the fraction of the detector volume
that is active, and to the $^{76}$Ge isotopic fraction, respectively. The
efficiency $\varepsilon_{fep}$ corresponds to the fraction of events that
deposit their entire energy at \qbb\ inside the active volume without
bremsstrahlung loss. Finally, $\varepsilon_{psd}$ represents
the efficiency of the signal acceptance by pulse shape discrimination (PSD).

The number of background events $\lambda_B$ in the region of interest (ROI)
around \qbb\ scales in a first approximation with the detector mass. It can be
expressed as follows:
\begin{equation}
     \label{eq:sensitivity2}
\lambda_B = M \cdot t \cdot BI \cdot \Delta{E}  \ \ .
\end{equation}
$BI$ is the background index for the ROI around \qbb\ in units of \ctsper\ and
$\Delta{E}$ is the width of the search window which depends on the energy
resolution at \qbb.

If the experiment can be carried out background-free, the sensitivity on the
half life scales with $M \cdot t$. In case of a sizable background
contribution ($\lambda_B>>1$), its statistical fluctuation can be assumed to
be Gaussian and the sensitivity would scale approximately with $\sqrt{(M \cdot
  t)/(\Delta{E}\cdot BI)}$.

\gerda\ has been conceived to proceed in different phases in order to fulfill
a quasi background-free condition in each of them. A sensitivity scaling
almost linearly with the exposure of the experiment is aimed for.

In Phase~I semi-coaxial Ge detectors from the Heidel\-berg-Moscow
(\hdm)~\cite{bib:HdM01} experiment and the International Germanium Experiment
(IGEX)~\cite{bib:IGEX02} were deployed after their
refurbishment. A background level of an order of magnitude lower than in those
former experiments was achieved within Phase~I~\cite{bib:GER13-0nbb}.

\begin{table*}[t]
\begin{center}
\caption{\label{tab:Ge_isotop} 
  Measured isotopic composition and calculated density of the
  $^{76}$Ge enriched germanium for \gerda\ Phase~II BEGe detectors. The reported
  measurements were performed via electron ionization and thermal ionization
  mass spectrometry (EI-MS, TI-MS) at ECP in Zelenogorsk, Russia, via neutron
  activation (k0NAAA) at IRMM in Geel, Belgium, and by means of inductively
  coupled plasma mass spectrometry (ICP-MS) at LNGS in Assergi, Italy. For
  comparison, the isotopic composition and density of natural germanium and of
  the \gerda\ $^{76}$Ge depleted BEGe detectors were added.
}
\begin{tabular}{ccllllll}
\hline
technique & Ref. &	~$^{70}$Ge	& 	~$^{72}$Ge	&	~$^{73}$Ge	&	~$^{74}$Ge	&	~$^{76}$Ge	& calculated density 		\\
 	& &	  			&	 			&	 			&	  			&	 			& [g/cm$^{3}$] \\
\hline
EI-MS, TI-MS&~$^\star)$ &	0.0002(1) 	&	0.0007(2)	&	0.0016(1)	&	0.1234(33) 	&	0.8742(36)	& 5.540(5)\\
k0NAAA  & ~\cite{bib:enrGe_k0naaa} 	&	0.001(1)	&	0.001(1)	&	0.001(1)	&	0.130(2) 	&	0.867(11)	& 5.539(11)\\
ICP-MS  & ~$^\star)$	&	0.0014(1) 	&	0.0003(5)	&	0.0011(10)	&	0.1065(141) &	0.8921(141)	& 5.550(20)\\
{\bf average}&	&0.001(1)	&	0.001(1)	&	0.001(1)	&	0.120(12) 	&	0.877(13)	& 5.540(18)\\
\hline
\genat 	& ~\cite{bib:natGe-composition} &	0.204(2)	&	0.273(3)	&	0.078(1)	&	0.367(2)	&	0.078(1)	& 5.323(4)	 \\
\gedep\ BEGe 	& ~\cite{bib:deplBEGe13} &0.223(8)&0.300(4)&0.083(2)&0.388(6)&0.006(2)&5.303(11)\\
\hline
\end{tabular}
\end{center}
$^\star)$ measured by \gerda\ and/or ECP
\end{table*}

For \gerda\ Phase~II another factor of ten in background reduction is
envisioned. This can only be achieved with an optimized experimental design
with particular care for the detectors. After several years of R$\&$D, a
customized version of the broad energy germanium (BEGe)
detector~\cite{bib:bege_pulse_shape} from Canberra with a thick entrance
window has been selected. The key to the superior rejection of background of
these detectors lies in the simple and powerful analysis of the digitized
waveform of the detector signals. In addition, external background events are
either fully absorbed in LAr or can be largely rejected on an event-by-event
basis by detecting scintillation light produced via interactions in
LAr~\cite{bib:tipp}.

This paper documents the entire production process from
 the enrichment (section~\ref{sec:enrichment}),
 the purification (section~\ref{sec:purification}),
 the crystal growth (section~\ref{sec:crystal_pulling})
 to the diode fabrication (section~\ref{sec:diode_production}).
 The precautions applied during the BEGe detector production in order to
 reduce cosmogenic-induced radioisotopes are described in
 section~\ref{sec:exposure}.

In total, 30 BEGe detectors were produced from newly acquired enriched
material with a $^{76}$Ge fraction of about 88\%. The detectors were produced
in two bat\-ches. The first one comprises seven detectors which were named
GD32(A-D) and GD35(A-C). After their fabrication and characterization in
vacuum cryostats (section~\ref{sec:performance_vacuum}) five of them were
deployed in \gerda\ during the data acquisition period of Phase~I. Their
performance in LAr is discussed in section~\ref{sec:performance_lar}.

The performance of all 30 BEGe detectors operated in vacuum cryostats
including an intercomparison and Monte Carlo (MC) studies will be presented in
an upcoming publication.

\section{Production of BEGe detectors for {\mbox{{\textsc GERDA}}} Phase~II}
\label{sec:production-intro}

\gerda\ is a \onbb\ experiment aiming for a quasi back\-ground-free ROI.
Therefore, highest resolution of the detectors and radiopurity of the entire
setup are of paramount importance. Within the germanium detector types, the
BEGe family combines advantageously high resolution and pulse shape
discrimination possibilities (see section~\ref{sec:bege-design}).  The costs
for enrichment of germanium in $^{76}$Ge from its natural abundance of 7.8\%
to about 88\% are compensated for by those for the reduced number of detectors
needed inclusive associated electronics.  For full depletion of BEGe detectors
of height of a few cm, adequate impurity levels of $\lesssim10^{11}$ are
needed to keep reverse bias voltages below 4\,kV. This requires a proper
purification before germanium crystals can be grown and converted into
operational diodes.

\subsection{Enrichment of $^{76}$Ge}
\label{sec:enrichment}

\begin{figure}[t]
\begin{center}
  \includegraphics[width=\columnwidth]{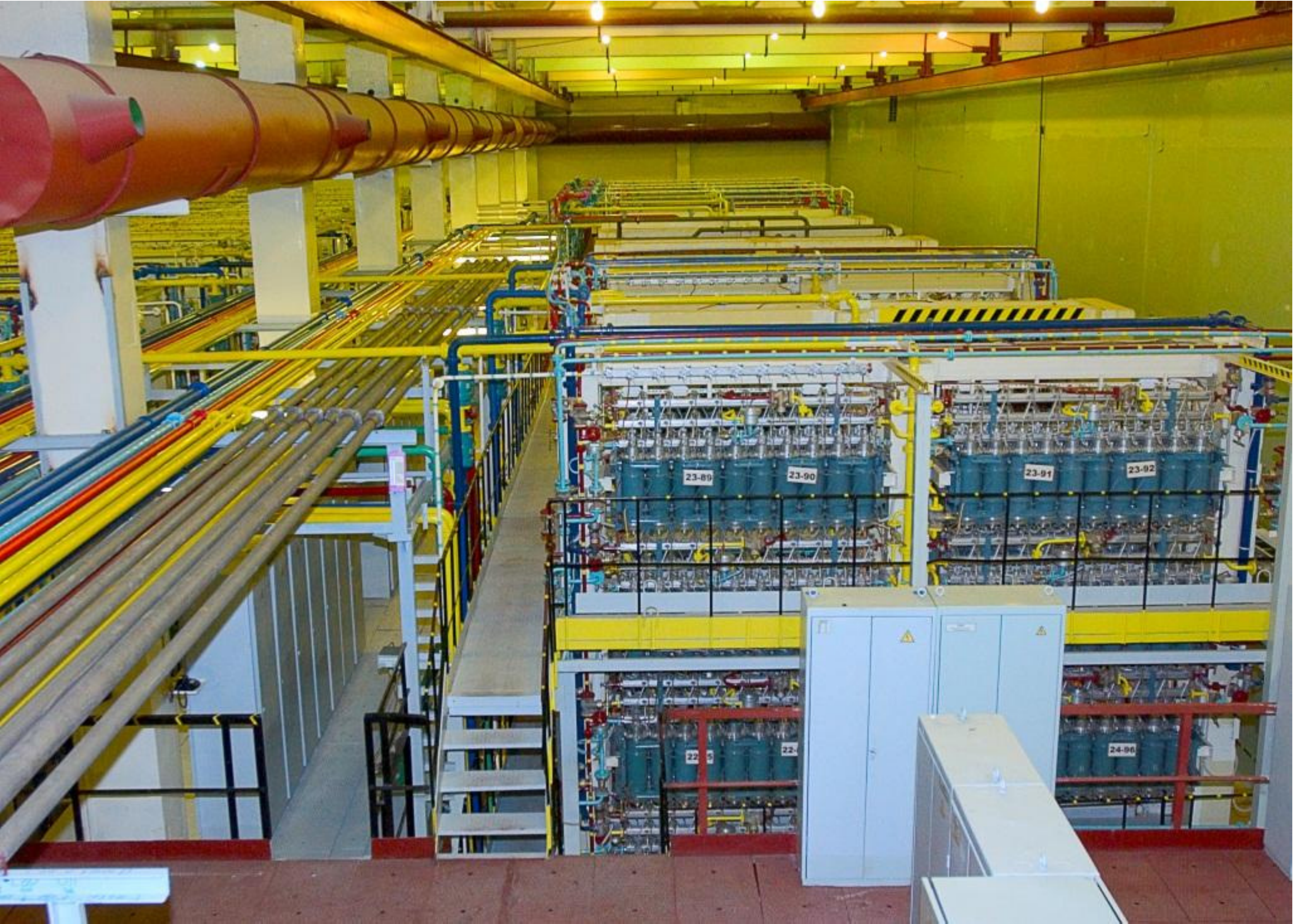}
  \caption{  \label{fig:ECP-centrifuges}
          Photo of a small part of the gas centrifuge system at Svetlana
          department of ECP. Photo courtesy of ECP.
}
\end{center}
\end{figure}

The enrichment in $^{76}$Ge for \gerda\ Phase~II detectors was performed at
the Svetlana Department of the Joint Stock Company ``Production Association
Electrochemical Plant" (ECP) in Zelenogorsk, Russia~\cite{bib:ecp_ru}.  Since
it is possible to bind Ge in gaseous GeF$_4$ compounds which possess a
relatively low vapor pressure at room temperature, the gas centrifuge
technique can be applied. The overall procedure is the following:
\begin{enumerate}
\item \genat\ fluorination: \genat\  $\rightarrow$ \genat F$_4$,
\item centrifugation process: \genat F$_4$ $\rightarrow$ \geenr F$_4$,
\item hydrolysis 
           within balloons: \geenr F$_4$ $\rightarrow$ \geenr
  O$_2$,
\item drying and calcination of \geenr O$_2$.
\end{enumerate}
Herein, \genat\ corresponds to natural germanium; \geenr\ stands for $^{76}$Ge
enriched germanium, which in the following will also be referred to as
`enriched'.  The gas centrifuge processing at ECP involves a large number of
centrifuges in series and parallel formations. A photo of one cascade of the
gas centrifuge assembly is shown in Fig.~\ref{fig:ECP-centrifuges}.

The annual productivity of the Svetlana Department facility is about
(80-100)\,kg of germanium at $\sim$88\% enrichment in $^{76}$Ge. The
production of the enriched germanium for the \gerda\ Phase~II BEGe detectors
started at the end of February 2005 and finished at the beginning of September
2005. In total, 53.4\,kg of GeO$_2$ powder was produced, which corresponds to
37.5\,kg of germanium enriched in
$^{76}$Ge~\cite{bib:enrGe-isotopic-comp}. After enrichment several subsamples
were measured for their isotopic abundances with different techniques. A
summary of the results is given in Table~\ref{tab:Ge_isotop}. The table also
includes the expected density of the final Ge crystals resulting from the
isotopic compositions under the assumption of a pure face-centered cubic
lattice. The density of enriched germanium remnants after crystal growth was
measured by \gerda\ and resulted in an average value of (5.552
$\pm$0.003(stat.)  $\pm$0.007(syst.))\,g$\cdot$cm$^{-3}$ at room
temperature. This result is in very good agreement with expectations.

In addition to the enriched material, ECP provided 49.2\,kg of GeO$_2$
depleted in \gesix\ (\gedep ; short form: `depleted'). Its isotopic
composition and calculated density are reported in
Table~\ref{tab:Ge_isotop}. The depleted germanium was a by-product of the
enrichment process and thus underwent the same chemical processing. For that
reason it was purchased for testing the following purification via
zone-refinement. Moreover, 34\,kg of additional depleted material was
purchased to test the full detector production chain including crystal growth
and diode production and to develop characterization procedures. For further
details see Ref.~\cite{bib:deplBEGe13}.

\subsection{Reduction and purification of germanium}
\label{sec:purification}

The ECP plant typically delivers $^{76}$GeO$_2$ powder of technical grade
quality which corresponds to 99.8\% purity level~\cite{bib:ecp_ru}. The
quality depends on the purity of the initial samples of \genat\ and \genat
F$_4$, as well as on the purity during the chemical transformation of \geenr
F$_4$ to \geenr O$_2$. For the construction of germanium diodes, however,
germanium of electronic grade; i.e., 99.9999\% purity (6N), has to be
available before the start of crystal growth.

A first step in increasing the purity of the germanium was accomplished
directly by ECP: improvements of clean conditions at work places, use of
de-ionized water for hydrolysis etc. led to a 99.99\% purity level (4N). This
was certified by three different Russian laboratories: the Central Laboratory
of ECP, the Analytic Certification Testing Center of the Institute of
Microelectronics Technology \& High Purity
Materials~\cite{bib:certCenterIMTHPM}, and the Certification Center of
Giredmet~\cite{bib:giredmet}.

For the transportation from Russia to Central Europe, the produced portions of
enriched and depleted GeO$_2$ were filled into plastic bags of about 1\,kg
each. Before the start of the purification process of the \geenr O$_2$, the
material was stored in the HADES underground laboratory in Mol, Belgium, from
April 2006 until March 2010.

\begin{figure}[t]
\begin{center}
  \includegraphics[width=\columnwidth]{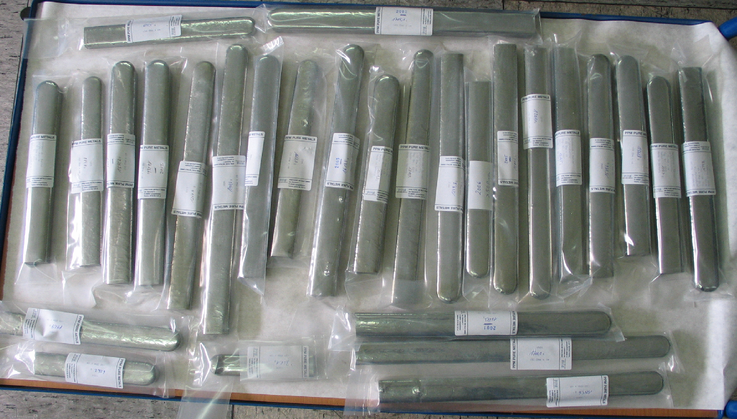}
  \caption{  \label{fig:ZR_Ge}
      Final inventory after the zone-refinement of the enriched germanium used
      for the production of the \gerda\ Phase~II detectors.
}
\end{center}
\end{figure}

For further purification, the company PPM Pure Metals
GmbH~\cite{bib:langelsheim} in Langelsheim, Germany, was selected. The overall
procedure at PPM was the following:
\begin{enumerate}
\item GeO$_2$ reduction: the GeO$_2$ powder was reduced in H$_2$ atmosphere to
  metallic Ge. The resulting metal ingots were cleaned and etched.
\item The Ge metal ingots underwent zone-refinement (ZR). The zone-refined
  bars were etched before they were packed in plastic bags (see
  Fig.~\ref{fig:ZR_Ge}) and delivered for the next production steps.
\end{enumerate} 

\begin{figure*}[t]
\begin{center}
\includegraphics[width=.99\columnwidth]{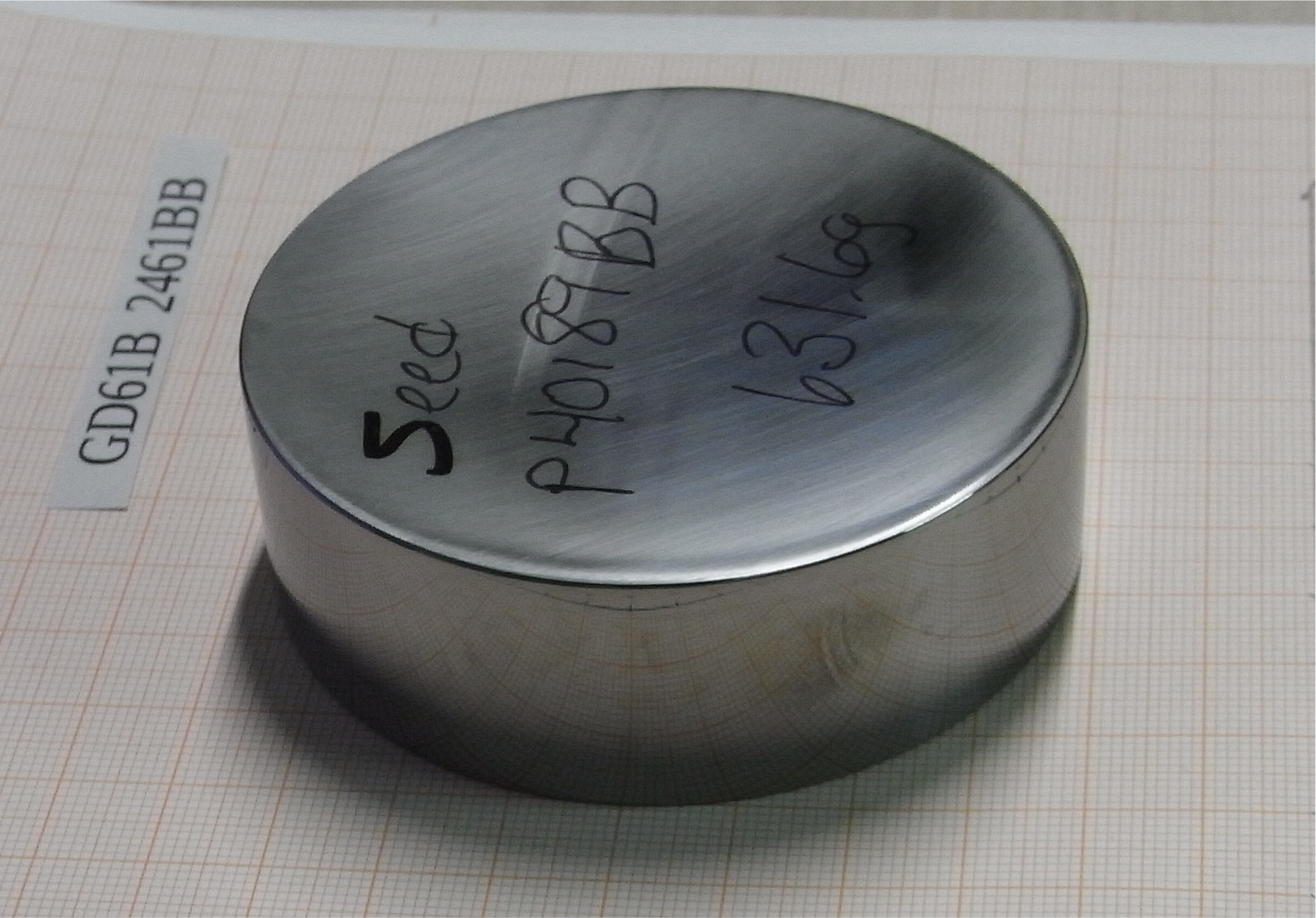}
\hfill
\includegraphics[width=.95\columnwidth]{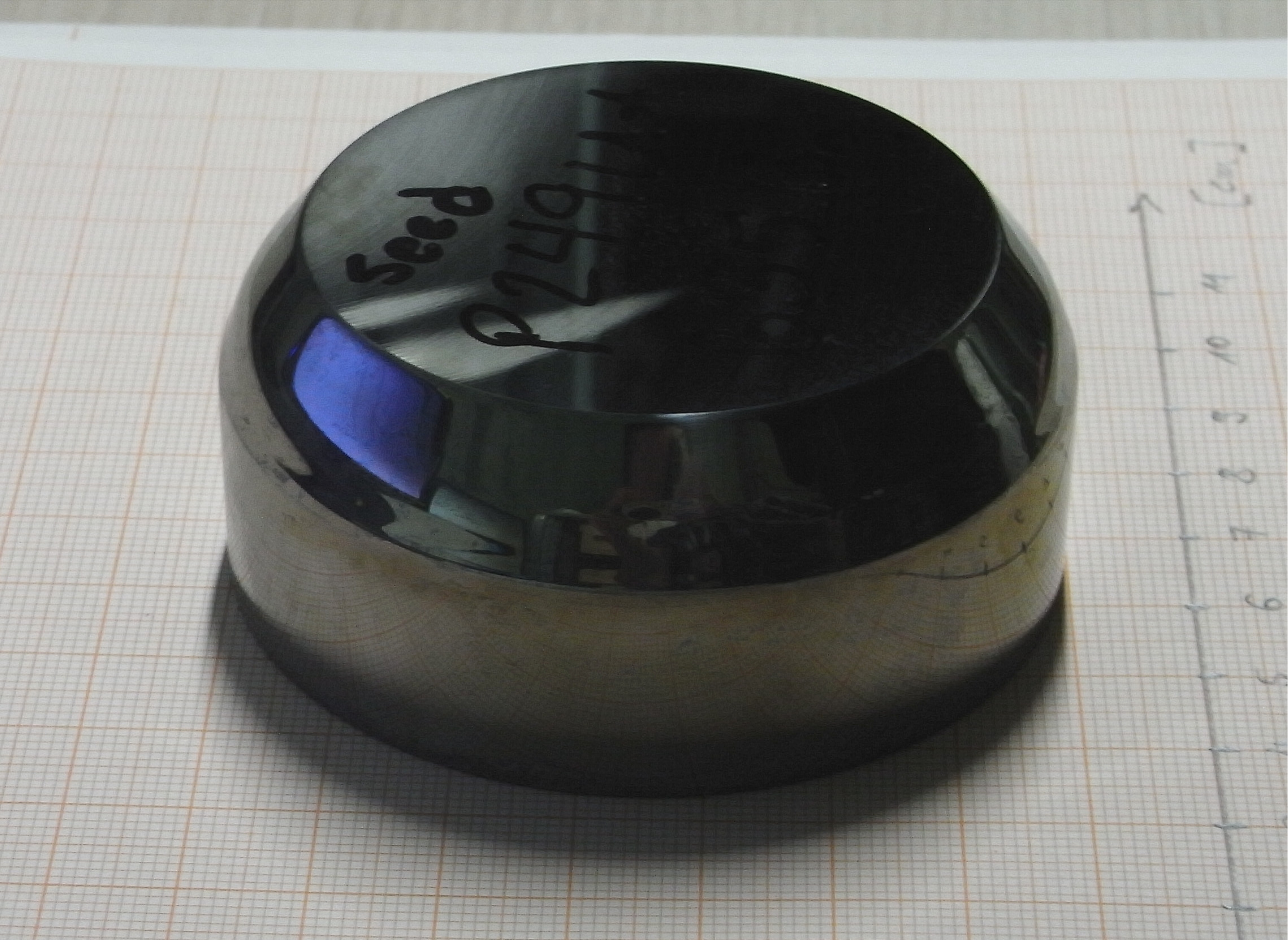}
\caption{\label{fig:2461BB-2491AA}
          Examples of a cylindrical (GD61B, left) and a conical (GD91A, right)
          crystal slice that were used for the production of \gerda\ Phase~II
          BEGe detectors.
}
\end{center}
\end{figure*} 

The achievable purification level, the mass yield and the potential change in
the isotopic composition were first tested with \gedep O$_2$. After reduction
and during ZR, measurements on extracted germanium subsamples were performed
by means of inductively coupled plasma mass spectrometry (ICP-MS), spark
source mass spectrometry (SS-MS) and resistivity measurements. 6N purity
electronic grade germanium has an intrinsic resistivity of about
50\,$\Omega\cdot$cm. The low resistivity tails ($<$50\,$\Omega\cdot$cm) of the
ingots were cut off and returned into the ZR furnace for 1-2 additional zone
refining steps. The last remaining tail is typically too small to be
reprocessed. The main conclusions from the experiences~\cite{bib:deplBEGe13}
with the {\gedep}O$_2$ were:
\begin{enumerate}
\item The purity of the Ge metal after reduction is the same as for the
  initial \gedep O$_2$. There was no sign that impurities were introduced.
\item No isotopic change was observed at the level of the measurement accuracy
  of $\pm$0.01\%.
\item The mass of the tail amounted to $<$1\,kg which translates in
   this case to a mass yield of $>$90\% for 6N material. 
\end{enumerate}
The following purification of the \geenr O$_2$ in spring 2010 went
smoothly. The purity of the \geenr O$_2$ material was the same as for the
\gedep O$_2$. An overall mass yield of 6N germanium of 94.5\% was
obtained. Combined with the residual low resistivity tail, 97.7\% of the
original 37.5\,kg of enriched germanium was finally available.

\subsection{Crystal growth}
\label{sec:crystal_pulling}

For further zone refinement and crystal growth the 35.5\,kg of 6N purified
enriched germanium was sent to Canberra Industries
Inc.~\cite{bib:canberra_usa}, Oak Ridge (TN), USA.

The enriched germanium was further zone-refined to 11N material. Then crystal
ingots with net carrier concentrations corresponding to 12N purity and with
specified dimensions and crystal dislocation densities \cite{bib:pureGeXtal}
were grown.

The crystal ingots were produced and delivered in two batches.  The first two
crystal ingots were grown in autumn 2011. Out of these seven crystal slices
were cut according to an optimized production scheme which was developed by
\gerda\ and Canberra during the depleted BEGe
production~\cite{bib:deplBEGe13}. After diode conversion (see
section~\ref{sec:diode_production}) and testing of the reliability of these
first prototypes in spring 2012, seven more ingots were grown. In total, 23
additional crystal slices were obtained.  Optimizing the mass yield was the
main goal when selecting the actual cut; however, in general \gerda\ and
Canberra aimed for specifications amongst them a diameter of 75\,mm with a
tolerance of $\pm$5\,mm, and a height of 30\,mm with a tolerance of
$^{+10}_{-5}$\,mm. All but three crystal slices met these specifications. In
two cases the diameters were 68.9 and 66.4\,mm, and in one other case the
height was 23.3\,mm. The average diameter and height of all crystal slices was
73.3 and 29.7\,mm, respectively, with ranges of about $\pm$3\,mm.

In optimizing the crystal slicing, conical tails and seed ends of ingots were
also considered. As a result, 21 crystal slices are cylindrical, whereas nine
are conical. Examples of a cylindrical and a conical crystal slice are
depicted in Fig.~\ref{fig:2461BB-2491AA}. The combined mass of all crystal
slices amounts to 20.8\,kg. Approximately 8.8\,kg of the loss was attributed
to seed-end and tail-end crystal parts and crystal remainders, whereas 5.5\,kg
of kerf; i.e., a mixture of germanium shavings, water and lubricant, was
collected in the process of grinding and lapping. All materials will be
prepared for recycling, i.e. a further cycle of chemical purification and zone
refinement before a new crystal growth.

All crystal slices for detector production and the crystal remainders were
shipped back to Belgium and stored in the HADES underground laboratory (see
section~\ref{sec:counteractions-exposure}) until the beginning of diode
conversion.

\subsection{Diode production}
\label{sec:diode_production}

The conversion of the germanium crystal slices into operational BEGe detectors
was performed at Canberra Semiconductors N.V.~\cite{bib:canberra_olen}, Olen,
Belgium, since detailed procedures had been developed in collaboration
before. The functionality and properties of BEGe detectors will be discussed
in sections~\ref{sec:bege-design} and~\ref{sec:bege-general-description}.

Three diodes were usually produced per week. In the case of 29 diodes 13\,g of
the original crystal mass was lost on average due to the groove
fabrication. In the case of detector GD76B, however, a larger crystal fraction
of about 370\,g (groove included) had to be removed to cure a micro-rupture
situated a few mm under the surface.

\begin{figure}[b]
\begin{center}
\includegraphics[width=.95\columnwidth]{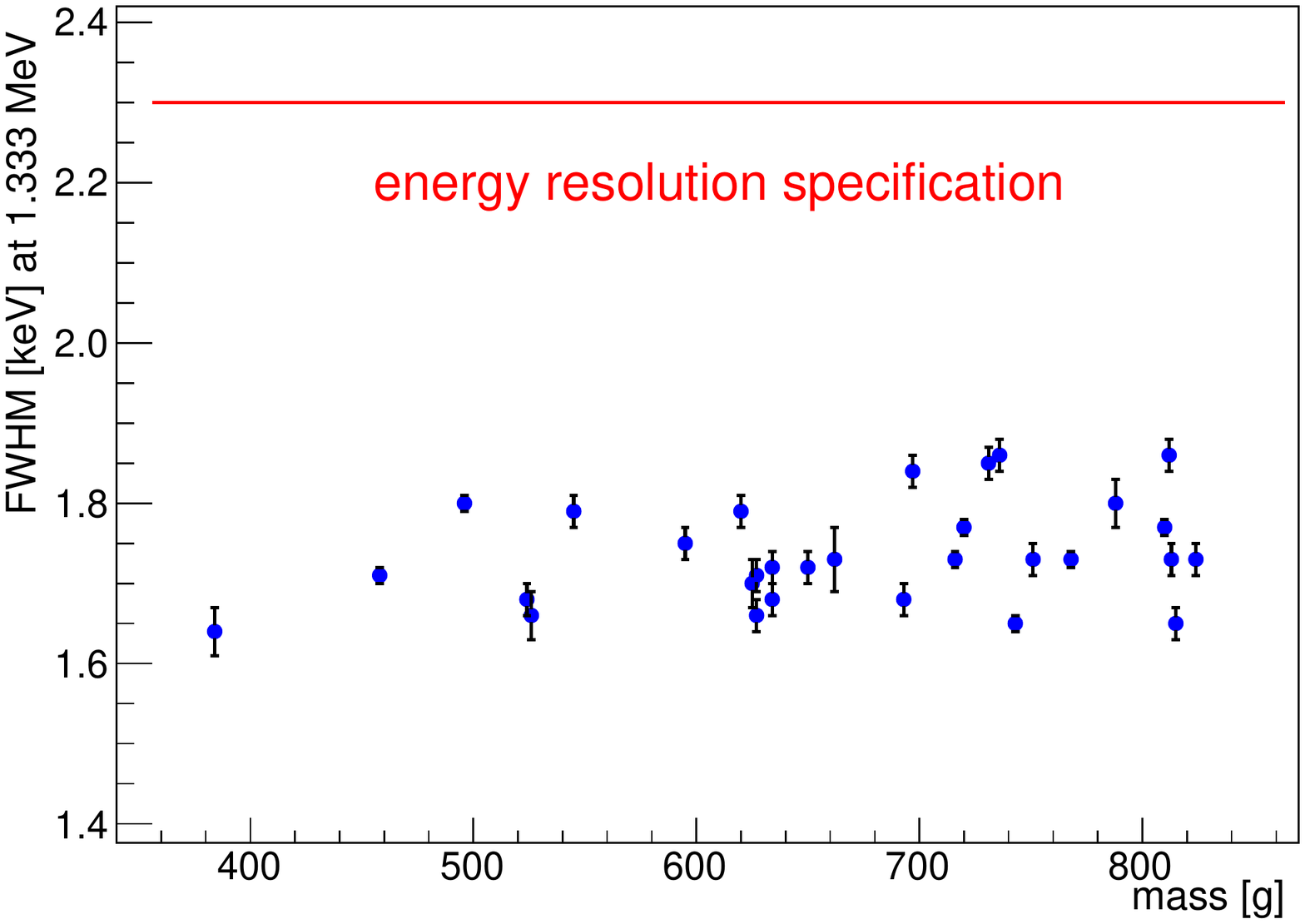}
\caption{\label{fig:eneres-vs-mass}
     Energy resolution of all 30 \gerda\ Phase II BEGe detectors as a
     function of their masses. The energy resolutions are expressed as full
     widths at half maximum of the $^{60}$Co peak at 1333\,keV. The
     error bars shown include only the uncertainties from the fit of the
     peaks. These were fitted with a step-like background and a Gaussian peak
     function.
}
\end{center}
\end{figure}

Prior to the delivery of the detectors, Canberra tested the diodes for their
basic parameters such as energy resolution, depletion voltage and leakage
current as a function of the applied high voltage. The following requirements
had to be met:
\begin{enumerate}
\item energy resolution: $<$\,2.3\,keV  full width at half maximum (FWHM)
  of the 1333\,keV $^{60}$Co $\gamma$-line,
\item operational (stable) voltage: $\le$\,4\,kV,
\item leakage current: $<$\,50\,pA at depletion voltage.
\end{enumerate}
Canberra was able to convert 29 out of 30 crystal slices into working
detectors fulfilling all three criteria. The energy resolution for all 30
detectors is illustrated as a function of the detector mass in
Fig.~\ref{fig:eneres-vs-mass}. The mean value of the energy resolution is
(1.74$\pm$0.07)\,keV at 1333\,keV. More details about the energy resolution,
depletion voltage, active volume and other spectroscopic properties of a
detector subset will be presented in section~\ref{sec:test-vacuum_results}.

One crystal slice (GD02D) turned out to have a non satisfactory impurity
distribution. This detector does not reach full depletion and the
corresponding voltage plateau; therefore it has a deteriorated charge
collection efficiency in some parts of the crystal. Nonetheless, this detector
will be deployed in \gerda\ Phase~II; its full or partial inclusion into the
analysis can be decided later.

The production of \gerda\ Phase~II BEGe detectors was optimized to achieve the
highest possible mass transfer from the original enriched germanium to the
final detector configuration. Out of 53.4\,kg of GeO$_2$, containing 37.5\,kg
of elemental enriched germanium, 30 detectors with a total mass of 20.0\,kg were
fabricated. This corresponds to a mass yield of 53.3\%. As shown in
Table~\ref{tab:mass-yield} the largest loss of germanium material occurred by
cutting the crystal ingots and while grinding and lapping the crystal
slices. The remainder are planned to be reprocessed.
\begin{table}[b]
\begin{center}
\caption{\label{tab:mass-yield}
    Mass yield from enriched germanium to the final 30 BEGe detectors for
    \gerda\ Phase~II. The mass transfer fractions are given relative to the
    original enriched GeO$_2$ material (3rd column) and to the purified
    metallic Ge used for crystal growth (4th column).
}
\begin{tabular}{lcrr}
\hline
\multicolumn{1}{c}{~~~~~~~~germanium}   & mass 
    &\multicolumn{2}{c}{rel. fraction}	\\
{operation}			&	[kg]
    &\multicolumn{1}{c}{ 
  [\%]}	&\multicolumn{1}{c}{ [\%]}\\
\hline
Ge in GeO$_2$ after enrichment    & 37.5 & 100.0 & --\textcolor{white}{....}\\
purified Ge for crystal growth    & 35.5 &  94.1\ & 100.0\\
cut crystal slices                & 25.2 &  67.2\ &  71.0\\
grinded and lapped crystal slices & 20.8 &  55.5\ &  58.6\\
operational detectors             & 20.0 &  53.3\ &  56.3\\
\hline
\end{tabular}
\end{center}
\end{table}

\section{Cosmic-ray activation of {\mbox{\textsc{GERDA}}} Phase~II germanium}
\label{sec:exposure}

\subsection{Cosmogenic production of radioisotopes in germanium}
\label{sec:intro-cosmogenics}

At sea level, secondary cosmic-rays consisting of fast nucleons (98\%) as well
as muons and muon-induced secondary neutrons (2\%) produce long-lived
radionuclides in materials via spallation
reactions~\cite{hayakawa1969,Ziegler1996}. The subsequent decays of these
radioisotopes generated inside the materials represent a serious source of
background in rare-event physics experiments.

In germanium long-lived radioisotopes such as $^{68}$Ge ($T_{1/2}$=270.8\,d)
and $^{60}$Co ($T_{1/2}$=5.27\,yr) are of main concern. The estimation of
their production rate depends on the isotopic composition of the germanium,
but also on varying neutron and proton fluxes at given locations and given
times~\cite{Ziegler1996}. Even though the proton flux at sea level is only
about 3\% of the neutron flux, its contribution to the radioisotope production
is about 10\% due to the more efficient stopping of protons by ionizing
interactions. The interpolated cross sections and semi-empirical models add
further uncertainties. In literature the activation rates at sea level for the
two most prominent radioisotopes $^{68}$Ge and $^{60}$Co in enriched germanium
(assuming 86\% $^{76}$Ge and 14\% $^{74}$Ge) vary between (1.0-13) and
(1.6-6.7)\,nuclei/(d$\cdot$kg), respectively~\cite{cebrian2010}. For the
following considerations the production rates of $^{68}$Ge and $^{60}$Co
reported in Table~\ref{tab:cosmic-production-rate-in-different-Ge-isotopes}
are applied. They are based on the excitation functions generated with the
SHIELD code~\cite{barabanov2006}. Regarding the enriched germanium used in
\gerda\ Phase~II detectors, the respective $^{68}$Ge and $^{60}$Co activation
rates at sea level are 5.8 and 3.3\,nuclei/(d$\cdot$kg). In these cases,
saturation at sea level is reached at $\sim$2300 $^{68}$Ge and $\sim$9200
$^{60}$Co nuclei/kg.

The contribution of the decays of $^{68}$Ge and $^{60}$Co radioisotopes to the
$BI$ of \gerda\ Phase~II detectors has been evaluated by means of MC
simulations~\cite{bib:GER13-bckg,thesis_matteo}.  The exponentially decreasing
background was averaged over the first three years of data collection in
\gerda\ Phase~II.  The $^{68}$Ge and $^{60}$Co radioisotopes would contribute
with 3.7$\times$10$^{-3}$ and 8.4$\times$10$^{-4}$\,\ctsper\, to the $BI$
around \qbb\ assuming 100 nuclei per kg detector mass for each of them.  A PSD
analysis (see sec\-tion~\ref{sec:vacuum_psa}) by itself can further reduce
this to 1.8$\times$10$^{-4}$ and 8.4$\times$10$^{-6}$\,\ctsper, respectively.

Without pulse shape analysis, however, $\sim$30 nuclei of $^{68}$Ge or
$\sim$120 nuclei of $^{60}$Co per kg germanium would already account for the
allowed background budget of 10${^{-3}}$\,\ctsper. Such concentrations in
$^{68}$Ge and $^{60}$Co are already reached after $\sim$5 and $\sim$36\,d of
exposure of unshielded enriched germanium at sea level, respectively. This
makes it mandatory to restrict the overall exposure to sea level cosmic
radiation during detector processing to a few days. As a consequence, large
efforts were made to minimize activation of the enriched germanium during the
entire production and characterization chain of the \gerda\ Phase~II BEGe
detectors.
\begin{table}[t]
\begin{center}
\caption{\label{tab:cosmic-production-rate-in-different-Ge-isotopes}
              Activation rates in nuclei/(d$\cdot$kg) of cosmogenic-induced
              radionuclides $^{68}$Ge and $^{60}$Co in Ge isotopes,
              in natural and in enriched germanium according to
              Ref.~\cite{barabanov2006}. In the case of the Ge isotopes the
              statistical standard deviations are in the range of (0.5-11)\%.
}
\begin{tabular}{lcccc}
\hline
Ge isotope/			&\multicolumn{2}{c}{neutron-induced}
                                        &\multicolumn{2}{c}{proton-induced}\\
isotopic composition&$^{68}$Ge      &$^{60}$Co  		&$^{68}$Ge      &$^{60}$Co      \\
\hline
$^{70}$Ge 	& 264.22& 1.56	& 17.17	& 0.17	\\
$^{72}$Ge 	& 50.56	& 2.6	& 4.78	& 0.29	\\
$^{73}$Ge  	& 25.44	& 2.8	& 2.54	& 0.34	\\
$^{74}$Ge	& 13.05	& 2.97	& 1.48	& 0.38	\\
$^{76}$Ge	& 3.68	& 2.85	& 0.54	& 0.46	\\
\hline
\genat  	& 74.84	& 2.56	& 5.60	& 0.32	\\
\geenr 		& 5.13	& 2.86	& 0.68	& 0.45	\\
\hline
\end{tabular}
\end{center}
\end{table}

\subsection{Actions to minimize activation of germanium}
\label{sec:counteractions-exposure}

\paragraph{Active removal of $^{68}$Ge and $^{60}$Co during germanium
processing:} During the enrichment process the centrifugation of the germanium
fluoride compounds separates light nuclides from the heavy fraction containing
$^{76}$Ge. The lighter stable Ge isotopes, which have larger activation cross
sections for $^{68}$Ge compared to heavier Ge isotopes, are suppressed. As a
consequence, the $^{70}$Ge abundance was reduced by more than three orders of
magnitude from 20.54 to 0.01\% (see Table~\ref{tab:Ge_isotop}). Since
$^{70}$Ge has a $\sim$70 times higher $^{68}$Ge production rate at sea level
than $^{76}$Ge, a reduction of $^{70}$Ge during the enrichment of $^{76}$Ge
considerably reduces the $^{68}$Ge activation rate. In total, the production
rate of $^{68}$Ge in enriched germanium is decreased by a factor of $\sim$14
compared to natural germanium. The activation rate for $^{60}$Co in enriched
germanium is similar to the one of natural germanium, as there is no
significant dependence of the activation rate on the mass number of the
germanium nucleus.

During chemical purification, zone refinement, and during growth of germanium
monocrystals impurities of U and Th as well as $^{60}$Co are efficiently
removed. Note, however, that during these refinement steps $^{68}$Ge cannot be
separated.

\paragraph{Optimization of germanium processing steps:}  
A significant exposure to cosmic radiation occurs during unshielded processing
at the manufacturer sites.  \gerda\ cooperated with all the manufacturers to
optimize their standard procedures in terms of speeding up their processing
steps. In case of the enrichment process, a notable improvement was
achieved. After the successful centrifugation and separation of light from
heavy isotopes, the production of new $^{68}$Ge and $^{60}$Co nuclides starts
right away as portions of gaseous GeF$_4$ flow from the last stages of the
cascade to the receiving balloons. According to ECP standard technology, the
collection of a \gerda -sized batch of GeF$_4$ into balloons, the chemical
conversion of this compound to germanium dioxide and the drying and
calcination process would last 40\,d on average. Following an upgrade of the
production plant the average time of production of the \gerda\ portion of
enriched germanium above ground was reduced to 74\,h (3.1\,d).

\begin{table*}[t]
\begin{center}
\caption{\label{tab:sites}
    Underground locations close to the manufacturer sites that were
    selected for the production of \gerda\ Phase~II BEGe detectors. The
    shielding power is expressed in terms of meters of water equivalent (m
    w.e.).
}
\begin{tabular}{lllrr}
\hline
processing step 	&location 		&UG site 	  & shield 	& distance\\
			&			&		  & [m w.e.]	& [km]	\\
\hline
enrichment 		&ECP, Zelenogorsk, RU	&concrete bunker  & 5		&$<$1	\\
purification to 6N 	&Langelsheim, GER 	&Rammelsberg mine & 80		&10	\\
crystal growth 		&Oak Ridge, TN, USA 	&Cherokee cavern  & 50		&7	\\
diode production 	&Olen, BE  		&HADES 		  & 500		&30	\\
detector characterization  &Mol, BE  		&HADES 		  & 500		&30	\\
operation in \gerda	&Assergi, IT		&\gerda\ at LNGS  & 3500	&--	\\
\hline
\end{tabular}
\end{center}
\end{table*}

\paragraph{Storage on-site:} 
As stated before, the germanium material was processed at different sites
throughout above ground. At these processing sites the germanium was stored in
nearby shallow or deep underground (UG) locations whenever it was not needed
for processing. Table~\ref{tab:sites} summarizes the processing steps applied
during the production of the \gerda\ Phase~II BEGe detectors, the sites where
these steps were performed, and the respective UG sites. Moreover, the
approximate shielding powers in terms of meters of water-equivalent (m w.e.)
of the UG sites and their distances to the manufacturers' sites are also
given. The processing was planned in close cooperation between the
manufacturers and the on-site \gerda\ collaborators that were responsible for
almost daily transportation of the material between the UG and processing
sites.

\paragraph{Transport:}
The transport of the enriched germanium from one processing site to another
was arranged in containers by truck and ship. Transport by aircraft was
excluded {\it a priori} since the cosmic-ray exposure in the higher atmosphere
is larger by orders of magnitude compared to sea
level~\cite{cosmic_rays_atm}. Since the transportation times were substantial,
which would lead to an unacceptable $BI$ contribution in \gerda\ Phase~II, a
shielded transport container was designed and built.

For the transport of GeO$_2$ powder from Zelenogorsk to Munich in February and
March 2006 a protective cylindric steel container was
used~\cite{barabanov2006}. Its dimensions are 140\,cm(D)$\times$126.5\,cm(H)
with a total weight of 14.5\,t. Inside the container a cavity of
54\,cm(D)$\times$40\,cm(H) can accommodate all the germanium. The expected
reduction factor of germanium activation due to nuclear as well as muon
components from cosmic rays was in the range of 10 for $^{68}$Ge and 15 for
$^{60}$Co~\cite{barabanov2006}. To demonstrate the possibility of
transportation, a conveyance from Zelenogorsk to Munich was accomplished in
2005. The journey lasted 20\,d. Taking the effect of shielding into account
this corresponds to $\sim$2\,d of exposure at sea level.

\begin{figure}[b]
\begin{center}
\includegraphics[width=\columnwidth]{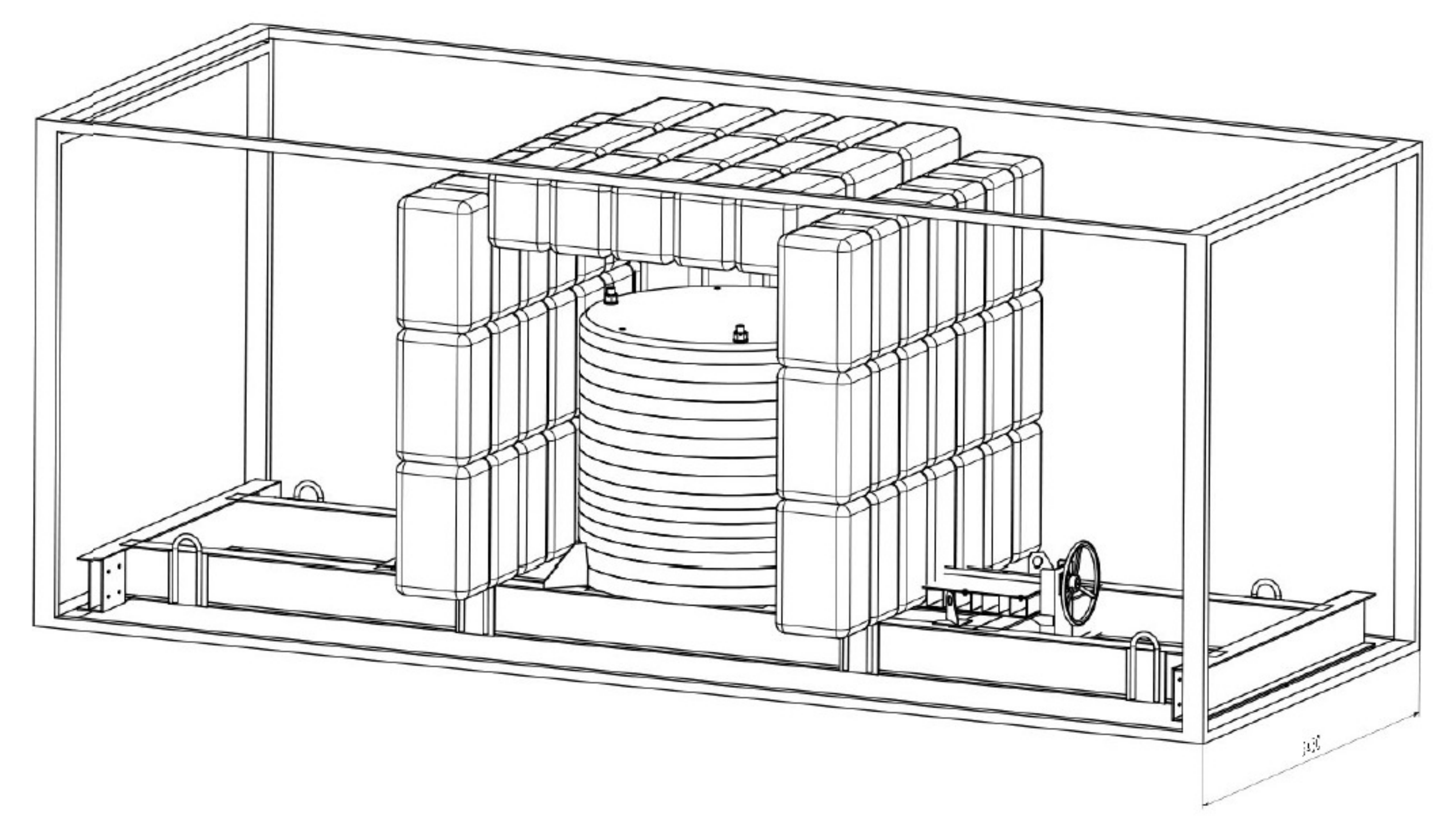}
\caption{\label{fig:shielding-container}
         Sketch of the shielded container used for the transport of the
         \gerda\ germanium material enriched in $^{76}$Ge. The shield
         has an inner iron container. This is surrounded by jerry cans
         filled with salt water and has a cavity to accommodate the germanium
         material. For visibility of the steel shield the front wall of the
         water jerry cans has been omitted.
}%
\end{center}%
\end{figure}

For further transports (including the return shipment from Europe to USA for
crystal growth) the container was upgraded. As shown in
Fig.~\ref{fig:shielding-container} the empty cavities between the container
roof and the steel shield were filled with jerry cans containing water with
30\,g salt per liter. The overall thickness of the water shield is 70\,cm
increasing the tare weight of the shielded container to 26\,t. According to
simulations~\cite{diplom_aaron}, both the $^{68}$Ge and $^{60}$Co production
rates are further reduced by about a factor of two.

\subsection{Tracking the exposure history and achieved activation levels}
\label{sec:full-tracking-exposure-history}

Starting from the enrichment phase, all periods between each processing and
transportation step above ground were documented in detail in a
database. Assuming a given production rate at sea level~\cite{barabanov2006}
and taking the transport container~\cite{barabanov2006,diplom_aaron} the
expected number of $^{68}$Ge and $^{60}$Co nuclei in each individual germanium
piece can be estimated for any given time.

\begin{figure}[t]
\begin{center}
\includegraphics[width=\columnwidth]{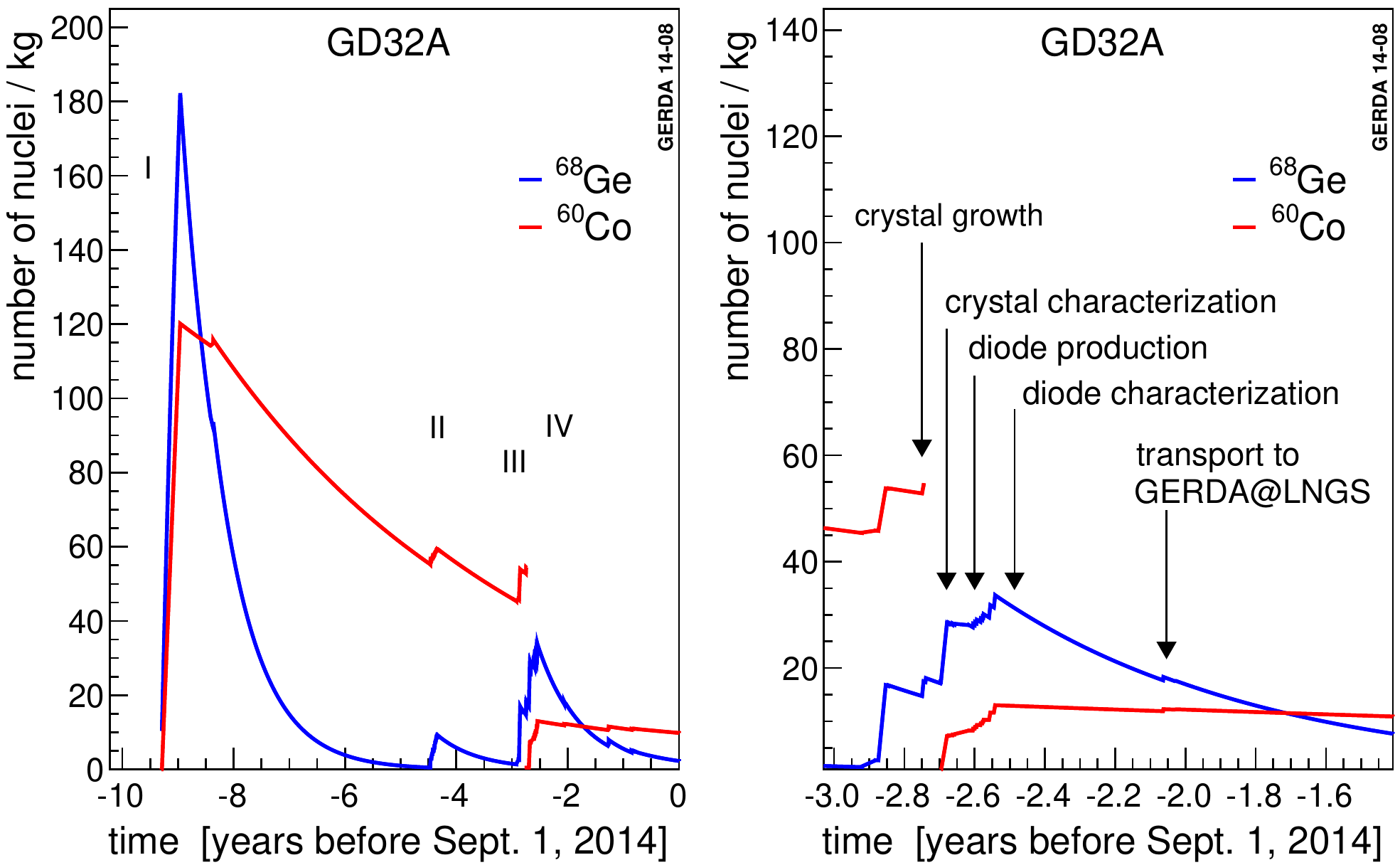}
\caption{\label{fig:activation_history}
       Left: Full history of the estimated number of cosmogenic-induced
       $^{68}$Ge and $^{60}$Co atoms in \geenr\ used for the manufacturing of
       detector GD32A. Major activation occurred during enrichment (I),
       purification (II), crystal growth (III) and diode conversion
       (IV). $^{60}$Co was removed completely at the time of crystal growth.
       Right: Zoom of the exposure history of detector GD32A during its
       crystal growth and diode conversion.
}
\end{center}
\end{figure}

Figure~\ref{fig:activation_history} depicts the history of the estimated
number of $^{68}$Ge and $^{60}$Co nuclei produced in one enriched BEGe detector
(GD32A). The activation history starting with the enrichment is shown up to
September 1, 2014. The individual processing steps and transport periods can
be clearly identified by the increase of the number of nuclei during the
unshielded times. Periods in which the germanium was shielded deep underground
become visible from the exponential decay of the shorter-lived $^{68}$Ge. The
exposure histories of the other enriched BEGe detectors are similar to that of
GD32A.

\begin{table}[b]
\begin{center}
\caption{\label{tab:activation_detectors}
        Number of $^{68}$Ge and $^{60}$Co nuclei in the seven enriched BEGe
        detectors as of September 1, 2014, according to their exposure
        histories and the production rates in Ref.~\cite{barabanov2006}. Due
        to the wide range of predicted production rates the reported numbers
        are quoted without uncertainties.
}
\begin{tabular}{lcccrc}
\hline
detector & mass &\multicolumn{2}{c}{$^{68}$Ge} &\multicolumn{2}{c}{$^{60}$Co}\\
	 &[kg]  & \# & \#/kg & \multicolumn{1}{c}{\#} & \#/kg \\
\hline
GD32A &	0.459 & 2.3 & 5.0 &  9.8 & 21.4	\\
GD35B &	0.717 & 4.1 & 5.8 & 18.6 & 26.0	\\
GD32C &	0.743 & 3.5 & 4.8 & 15.9 & 21.4	\\
GD32B &	0.723 & 3.6 & 4.9 & 15.3 & 21.1	\\
GD32D &	0.768 & 3.4 & 4.4 & 16.2 & 21.1	\\
GD35C &	0.812 & 3.1 & 3.9 & 13.6 & 16.8	\\
GD35A &	0.635 & 3.9 & 6.1 & 14.3 & 22.5	\\
\hline
Total & 4.857 & 23.9 & 4.9 &103.7& 21.4  \\
\hline
\end{tabular}
\end{center}
\end{table}
Table~\ref{tab:activation_detectors} summarizes the estimated amount of
$^{68}$Ge and $^{60}$Co, respectively, by September 1, 2014, for the 
subset of detectors delivered first. On average, $\sim$5 $^{68}$Ge and $\sim$21
$^{60}$Co atoms/kg are expected. According to background studies by
\gerda\ the decays of these radionuclides $^{68}$Ge and $^{60}$Co over a
period of three years of non interrupted data collection will lead in both
cases to a background rate of 1.8$\times$10$^{-4}$\,\ctsper\, at
\qbb~\cite{bib:GER13-bckg}. Taking advantage of the background rejection via
pulse shape analysis (see section~\ref{sec:bege-general-description}) the $BI$
contribution from the two radionuclides can be lowered to
9.1$\times$10$^{-6}$\,\ctsper\, and 1.8$\times$10$^{-6}$\,\ctsper,
respectively~\cite{thesis_matteo}. As a consequence, the background
contribution from $^{68}$Ge and $^{60}$Co decays should be at least 50 times
lower than the total $BI$ envisioned for \gerda\ Phase~II, even if the
production rates at sea level used in the current calculation were
underestimated by factor of two.


\section{Detector characterization in vacuum}
\label{sec:performance_vacuum}

After confirmation of proper functionality, stable operation and detector
parameters by the manufacturer, the \gerda\ collaboration performed
cross-checks (`acceptance tests') of the manufacturer specifications and
characterization tests. The goal was to characterize detector properties that
cannot be easily accessed by the manufacturer or by \gerda\ after integration
into the experiment.

Before discussing the measurements by \gerda\ and the results obtained for the
first seven enriched BEGe detectors in section~\ref{sec:test-vacuum_results},
some basic concepts of BEGe-type detectors are introduced in
sections~\ref{sec:bege-design} and~\ref{sec:bege-general-description}.

\subsection{Design of BEGe detectors}
\label{sec:bege-design}

\begin{figure}[b]
\begin{center} 
   \includegraphics[width=0.99\columnwidth]{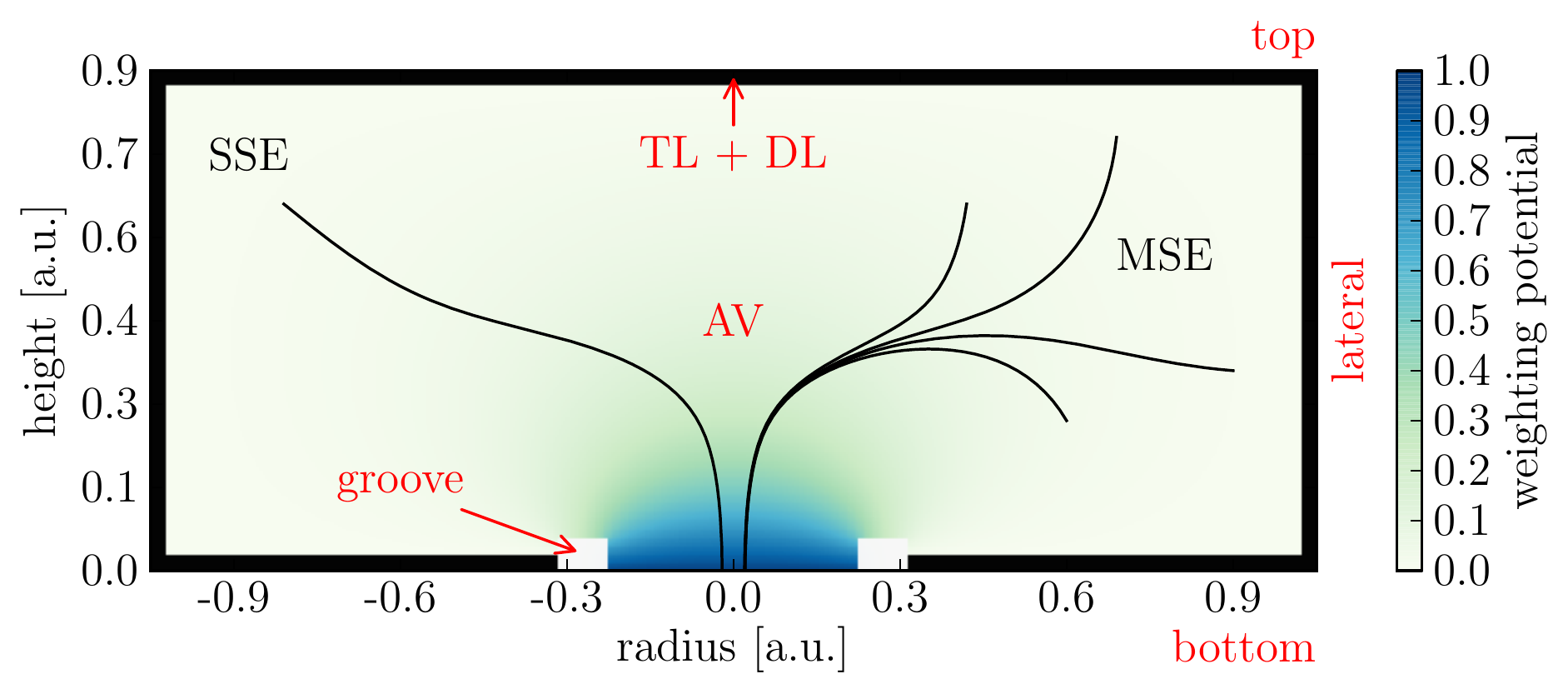}
   \caption{\label{fig:bege-profile}
              Generic view of a BEGe diode. The p+ read-out electrode (center,
              bottom) is separated from the n+ electrode by a groove covered
              by an insulating passivation layer. The n+ electrode consisting
              of an inactive dead (DL) and transition (TL) layer with reduced
              charge collection surrounds the active volume (AV). The
              dimensionless weighting potential is color-coded and strongest
              close to the p+ electrode. Examples of a single-site event (SSE)
              and of a multi-site event (MSE) are illustrated by means of the
              hole trajectories of the individual energy depositions.
}
 \end{center}
\end{figure}

\gerda\ has chosen a modified thick window Broad Energy Germanium (BEGe)
detector manufactured by Canberra as the detector type for Phase~II.
Compared to the semi-coaxial detectors used in \GERDA\ Phase~I, the BEGe
detector design shows smaller dimensions and thus smaller mass. Due to a
different layout of the electrodes (see Fig.~\ref{fig:bege-profile}) the
electric field profile in BEGe detectors differs strongly from the one in
semi-coaxial detectors.

The selected BEGe detectors are made of p-type germanium; they comprise a
`wrap around' n+ electrode known as `lithium dead layer', a p+ electrode
acting as electron blocking contact, and an intercontact insulating
surface. The first two items are achieved by lithium diffusion and boron
implantation. For the third item a small annular concentric groove between the
p+ and n+ electrodes is produced and covered by an insulating silicon monoxide
layer which is known as `passivation layer'. This layer helps to keep
steady-state currents (so-called `leakage currents') stable over time. The
currents are induced on the detector surface and in the bulk by some finite
conductivity in the reverse direction and should be kept at a few tens of pA.

In p-type BEGe detectors the dimensionless `weighting potential' $\Phi$ shown in
Fig.~\ref{fig:bege-profile} peaks strongly close to the central p+
electrode. Ionization will create electrons and holes which drift due to
the applied potential and the field created by the space charge of the
depleted diode. The time dependent induced current $I(t)$ on the p+ electrode
is given by the Ramo-Shockley theorem~\cite{bib:shockley-ramo-theorem} as:
\begin{equation}
 \label{eq:shockley-ramo_theorem} 
  I(t) = q\cdot \vec{v}(\vec{r}(t)) \cdot \nabla \Phi(\vec{r}(t))
\end{equation}
where $q$ stands for the drifting charge and $\vec{v}(\vec{r}(t))$ for the drift
velocity at position $\vec{r}(t)$. Holes drift to the p+ electrode along the
region around the central axis, irrespective of the starting point (`funnel'
effect); $I(t)$ peaks at the end of the drift where $\nabla\Phi$ is
largest. Hence, the maximum $A$ of $I(t)$ is directly proportional to the
deposited energy $E$. Electrons drift through volumes with low $\nabla \Phi$
and hardly contribute to $A$. That means that $A/E$ is constant for all
single-site events (SSE) except for ionizations in a small volume close to the
p+
electrode~\cite{bib:bege_pulse_shape,bib:deplBEGe13,bib:bege_pulse_simulation}.
In contrast, for multi-site events (MSE) the drift times of holes from several
simultaneous energy depositions are in general different and hence $A/E$ of
the summed signal is reduced. For ionizations in the n+ transition layer (like
from surface $\beta$-events) the diffusion time is comparable to the drift
time and hence $A/E$ is also reduced. For p+ surface events electrons drift
through the volume with largest $\nabla\Phi$, and hence $A/E$ is larger than
for SSE due to the increased displacement current.  The latter is also the
case for events close to the groove.

The layout imposes tighter constraints on the impurity and defect
concentrations in the starting crystal material that are more demanding to
achieve. These requirements arise due to the non-coaxial electrode arrangement
in BEGe detectors and the resulting electric field profile in order to achieve
complete charge collection within the detector volume.

\subsection{General properties of BEGe detectors}
\label{sec:bege-general-description}

\paragraph{Depletion voltage:}
The depletion voltage of a Ge detector is defined as the reverse bias voltage
that electrically fully depletes the diode of free charge carriers. It
strongly depends on the net impurity concentration and its gradient, on the
detector dimensions and on the read-out electrode sizes. The best performance
in terms of energy resolution is achieved at full depletion but it is
necessary to keep the leakage currents small. The depletion voltage is usually
determined via an irradiation of the detector with a $\gamma$-ray calibration
source and a stepwise increase of the voltage up to the value recommended by
the manufacturer. During this high voltage (HV) scan, detector operational
parameters such as the energy resolution $\Delta E$, the peak position and the
peak integral of prominent $\gamma$-lines are monitored. As soon as the
detector is electrically fully depleted, these parameters reach almost
constant plateaus. The measured curves are fitted. Then, the depletion
voltages are defined as those fit points, at which 99\% of the optimal $\Delta
E$ and maximum peak count rate as well as 99.9\% of the highest peak position
is obtained.  Note that a $\sim$1\,${}^{0\!}\!/\!_{00}$ drift of the peak
position corresponds to a $\sim$1\,keV shift in energy, a $\sim$1\% reduced
peak integral to a $\sim$1\% reduced active volume, and a $\sim$1\% lower
energy resolution to a $\sim$0.02 keV broader $\gamma$ peak -- all  at
$\sim$1\,MeV $\gamma$s detected by a BEGe detector.

\paragraph{Active volume:}
The p-type BEGe detectors (Fig.~\ref{fig:bege-profile}) have an internal
active volume (AV) in which the charge collection efficiency (CCE) is maximal
($\epsilon=1$). Gamma-rays fully absorbed in this volume contribute to the
full-energy peaks (FEP). The AV is surrounded by a transition layer (TL) with
reduced CCE ($0<\epsilon<1$) and a low electric
field~\cite{majorana2013}. Charges released in this region diffuse into the AV
only in part. Finally, the TL is covered by a thin conductive lithium-doped
layer in which the CCE is entirely suppressed ($\epsilon=0$) (see
section~\ref{sec:bege-design}) and is therefore called dead layer (DL). This
notation is more detailed because of the specific dependence of \gerda\ on the
\onbb\ signal and possible partial energy losses. Previously, in standard
$\gamma$-ray spectroscopy DL and TL were lumped together as a single totally
inactive `dead layer'.

A precise knowledge of the
de\-tec\-tor-specific active volume (AV) and its uncertainty is of great
importance for \gerda. There are two possibilities to determine it. Firstly,
high energy calibration sources can be used to irradiate directly the
AV. Secondly, low energy probes are used to measure the full-charge collection
depth (FCCD); i.e., the sum of TL and DL thicknesses. In this case, the AV
fraction $\factvol$ is deduced via a subtraction of the FCCD volume from the
detector volume which was calculated from the measured geometrical
dimensions. Complementary measurements are needed in order to reduce
systematic uncertainties. A difficulty in scanning is present since detectors
are housed in cryostats with thick end caps. This was overcome via a surface
scan with low energy $\gamma$-ray sources able to penetrate the cryostat
endcaps.

\paragraph{Pulse shapes:} 
Efficient
background suppression is of pa\-ra\-mount importance for \gerda. \onbb\
events in germanium are characterized by the absorption of two emitted
$\beta$-particles within a small volume of few mm$^3$ which is interpreted as
a SSE. On the contrary, $\gamma$-rays of similar energy can undergo multiple
Compton scattering leading to MSE. Based on these pulse shape differences
background events can be identified and suppressed.

In order to study the pulse shape discrimination (PSD) power, high energetic
$\gamma$-ray calibration sources are often used. Further, fine-grained surface
scans with collimated low energy $\gamma$-ray probes help in understanding local
differences in the pulse shape response and deduce detector intrinsic
properties such as the crystal lattice and electron and hole mobilities in
germanium. 

\subsection{Tests in vacuum cryostats}
\label{sec:test-vacuum_results}

All 30 \gerda\ Phase~II BEGe detectors were delivered in a Canberra dip stick
vacuum cryostat of type 7500SL with a 4'' endcap
diameter~\cite{bib:pst_7500_SL}. Characterization tests were performed in the
HADES underground research laboratory in Mol, Belgium, at 30\,km distance from
the diode manufacturer. Inside HADES an area of $\sim$14\,m$^2$ was equipped
with several static measurement tables and automatized movable scanning
setups, 33 radioactive sources, and two types of data acquisition systems:
Multi-Channel Analyzers (MCA) and 100\,MHz Struck Flash Analog Digital
Converters (FADC). The signals were read out with a charge sensitive
preamplifier provided by Canberra (model 2002CSL) and -- in case of the FADC
systems -- digitized. Then, the energy $E$ and the maximum of the current
pulse $A$ were reconstructed by digital signal processing using a
semi-Gaussian shaping. This offline analysis was performed with the software
tool \gelatio~\cite{bib:gelatio1} following the procedure described in
Ref.~\cite{bib:gelatio2}. Data storage systems and a network for remote
control and data transfer were installed. This infrastructure, called HEROICA,
had a screening capacity of two detectors per week in case only standard
measurements were performed. A more detailed description of the screening
facility can be found in Ref.~\cite{bib:Heroica}.

Most of the tests applied to the new enriched BEGe detectors are based on
campaigns and protocols of natural and depleted BEGe
detectors~\cite{bib:deplBEGe13} which served as prototypes to verify the
production chain and the detector performance compared to former detector
designs. In addition, several non-standard tests were applied on particular
detectors. In the case of the active volume (AV) determination, for instance,
many systematic effects were investigated. These included a remea\-surement of
the diode masses and dimensions, a cross-check of the diode position inside
the cryostat endcap, dead time uncertainty estimations, germanium density
measurements, and intercomparisons of source activities and of MC code
versions.

\subsubsection{Depletion voltage and energy resolution}
\label{sec:hv_scans}

The depletion voltage of the first seven enriched BEGe detectors was measured
by using pointlike $^{60}$Co sources and performing high voltage (HV) scans in
steps of (50-100)\,V typically from 500\,V up to the voltage $V_r^{C}$ of
several kV which was recommended by Canberra. At each intermediate voltage
point the energy spectrum was measured and three parameters were monitored:
the peak position (PP), the peak integral (PI) and the energy resolution
$\Delta E$ of the two $^{60}$Co $\gamma$-lines. $\Delta E$ is expressed in
terms of FWHM and was calculated via a fit function consisting of a Gaussian
distribution for the peaks and a step-like function describing the background
and Compton continua.

\begin{table*}[t]
\begin{center}
\caption{\label{tab:ene-res-depl-volt}
        Voltage parameters and energy resolutions of the first seven enriched
        BEGe detectors for \gerda . Measurements performed
        by the manufacturer Canberra and by \gerda\ are marked by `C' and
        `G', respectively.  All other abbreviations are explained in the
        text. The manufacturer results were provided without an uncertainty
        quotation. The \gerda\ voltage values have uncertainties of around
        $\pm$200\,V. Due to the systematic uncertainties the values for the
        energy resolution are
        rounded to the significant digits. 
}
\begin{tabular}{lccccc}
\hline
detector &\multicolumn{2}{c}{voltage}  &\multicolumn{3}{c}{energy resolution} \\
  &$V_{r}^{C}$; $V_{d,PP}^{C}$  &$V_{d,PI}^{G}$;$V_{d,PP}^{G}$;$V_{d,\Delta E}^{G}$ &$\Delta E^{C}$ [keV] &$\Delta E^{G}$ [keV] &$\Delta E^{G}$[keV] \\
	&[kV]	&[kV]	&at 1333\,keV&at 1333\,keV &at 2615\,keV\\
\hline
GD32A &3.0; 2.5 &2.1; 2.4; 2.6 &1.695 &1.73 &2.46 \\
GD32B &4.0; 3.5 &2.1; 2.7; 3.0 &1.747 &1.77 &2.49 \\
GD32C &4.0; 3.5 &2.9; 3.2; 3.7 &1.658 &1.70 &2.41 \\	
GD32D &4.0; 3.5 &2.2; 2.7; 2.8 &1.757 &1.65 &2.45 \\
GD35A &4.0; 3.0 &2.6; 2.6; 2.7 &1.785 &1.71 &2.40 \\
GD35B &4.0; 3.5 &2.5; 2.9; 3.5 &1.748 &1.80 &2.57 \\
GD35C &3.5; 3.0 &2.3; 3.0; 3.3 &1.643 &1.78 &2.50 \\
\hline
\end{tabular}
\end{center}
\end{table*}

The HV scan curves of the detectors GD32A and GD35B are depicted exemplarily
in Fig.~\ref{fig:GD32A-HVscan} and~\ref{fig:GD35B-HVscan}. The three curves of
a single detector converge to an almost constant value at approximately the
same voltage. Within this study the depletion voltages $V_{d,\Delta E}$,
$V_{d,PP}$, $V_{d,PI}$ were deduced, at which the single parameters reach an
almost constant value as described in
section~\ref{sec:bege-general-description}. Table~\ref{tab:ene-res-depl-volt}
summarizes the results obtained for the seven enriched BEGe detectors. In
particular, the following observations are made:
\begin{figure}[b]
\begin{center}
\includegraphics[width=0.9\columnwidth]{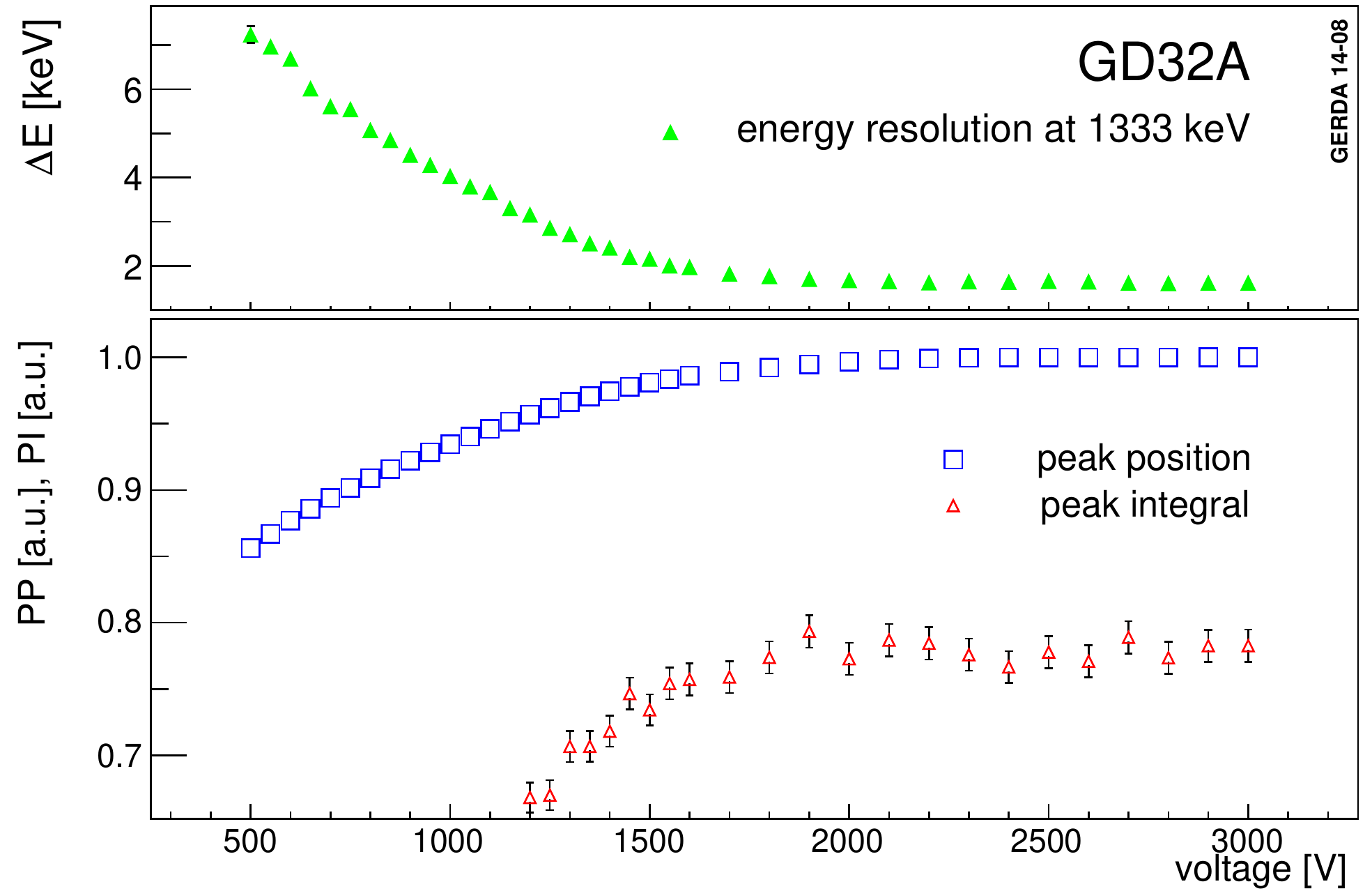}
\caption{\label{fig:GD32A-HVscan}
        $^{60}$Co HV scan of detector GD32A: energy resolution
        $\Delta E$ (as FWHM at 1333\,keV), peak position (PP) and peak
        integral (PI) as functions of the applied voltage.
} 
\end{center}
\end{figure}

\begin{figure}[b]
\begin{center}
\includegraphics[width=0.9\columnwidth]{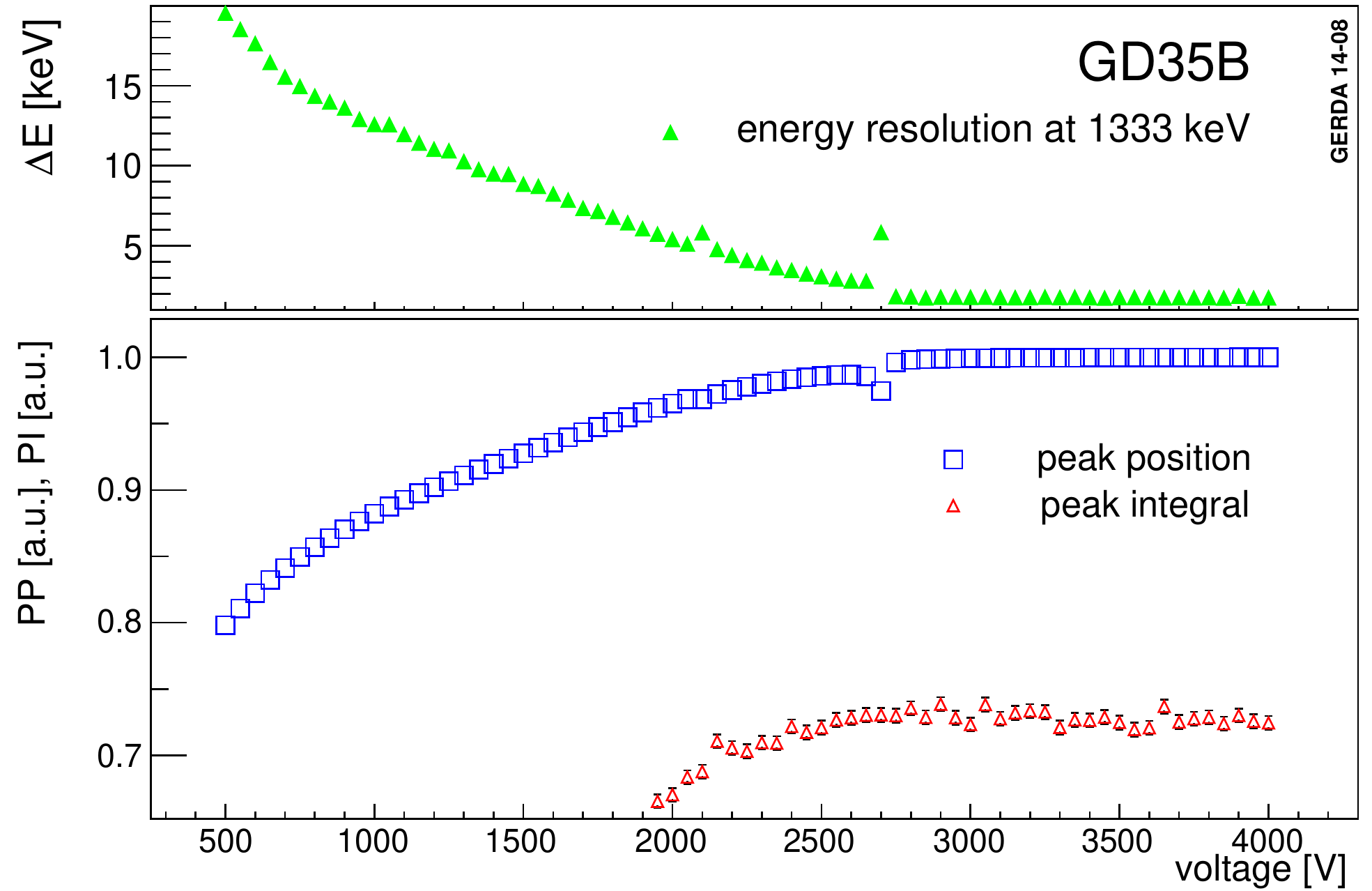}
\caption{\label{fig:GD35B-HVscan}
        Same as in Fig.~\ref{fig:GD32A-HVscan} for GD35B.
        A discontinuity around 2.7\,kV is visible which originates from the
        so-called  `bubble/pinch-off' effect.
}
\end{center}
\end{figure}

$(i)$  The measured reference voltages typically fulfill the relation
 $V_{d,PI}\lesssim V_{d,PP}\lesssim V_{d,\Delta E}$. The values $V_{d,PI}$,
 $V_{d,\Delta E}$ and $V_{d,PP}$ are compatible for both 1173\,keV and
 1333\,keV $\gamma$-lines.  The detectors are usually operated at the voltage
 $V_{r}^{C}$ recommended by the manufacturer, who measures typically only the
 peak position $V_{d,PP}^C$ vs. voltage. The value of $V_r^C$ is usually
 500\,V above the measured $V_{d,PP}^C$ value. Canberra's $V_{d,PP}^C$ values
 are slightly higher than the voltage region defined by \gerda\ in which the
 PP is above 99.9\% of its maximum value. This guarantees that the three
 detector parameters will always be in an optimum region for operation.

$(ii)$ The reference energy resolution $\Delta E$ of the seven BEGe detectors
  was deduced from a measurement in which they were operated at the
  recommended voltage $V_{r}^{C}$.

 The peaks were fitted with a Gaussian function and a step-like plus a
 constant background term. The uncertainties due to the fit are at the 10 eV
 level. The systematic uncertainty is estimated in this case to 0.03\,keV.
 All detectors have a similar $\Delta E$. For the $\gamma$-lines from
 $^{60}$Co and $^{208}$Tl decays at 1333 and 2615\,keV the averaged values are
 (1.73$\pm$0.05) and (2.47$\pm$0.05)\,keV, respectively.  All the $\Delta E$
 values are $\sim$30\% better than those of the \gerda\ semi-coaxial detectors
 operated in vacuum cryostats (see Ref.~\cite{bib:Marik-thesis}). In general,
 the \gerda\ $\Delta E$ values are in good agreement with the manufacturer's
 specifications.
 
$(iii)$ The parameter dependencies of all detectors follow expectations from the
  known impurity densities and geometries of the detectors. For detectors
  GD35A and GD35B (Fig.~\ref{fig:GD35B-HVscan}) the curves of all three
  parameters were found to be very similar. The diode and read-out electrode
  geometries of the two detectors are comparable, and the impurity
  concentrations -- as confirmed by the manufacturer -- are similar. Moreover,
  both detectors exhibit the so-called `bubble'~\cite{bib:bubble-effect} or
  `pinch-off' effect~\cite{bib:pinch-off-effect}: In a voltage interval of a
  few tens of volts just below the depletion voltage, an island in the central
  region forms in which the total electric field becomes zero. Depending on
  their starting position almost all charge clouds drifting to the read-out
  electrode might cross this island and get affected. This gives rise to the
  observed broader energy resolution and a peak position instability, which
  leads in the above two cases to a discontinuity around (2.3-2.7)\,kV.

\subsubsection{Active volume determination}
\label{sec:dl_av_determination}

The active volume fractions $\factvol$ of the seven enriched BEGe detectors
were determined by an intercomparison of ca\-li\-bra\-tion data with simulated
calibrations of the same experimental setup~\cite{bib:corrado}. Calibration
spectra were taken for two complementary types of $\gamma$-ray
sources. Firstly, uncollimated low-energy $\gamma$-ray emitting $^{241}$Am
sources were deployed 19.8\,cm away from the cryostat endcap to probe the FCCD
and thus the combined TL and DL thickness. The $\factvol$ fractions are
deduced indirectly by subtraction of the summed TL and DL volume from the
overall detector volume. Secondly, higher energy $^{60}$Co sources with an
activity of $\sim$(6-14)\,kBq were positioned at the same distance from the
cryostat endcaps as in the case of the $^{241}$Am sources. $^{60}$Co sources
with activities calibrated at a $\pm$1\% level were used to probe directly the
AV by irradiating the entire bulk of the detector diode.

\begin{figure}[t]
\begin{center}
\includegraphics[width=0.9\columnwidth]{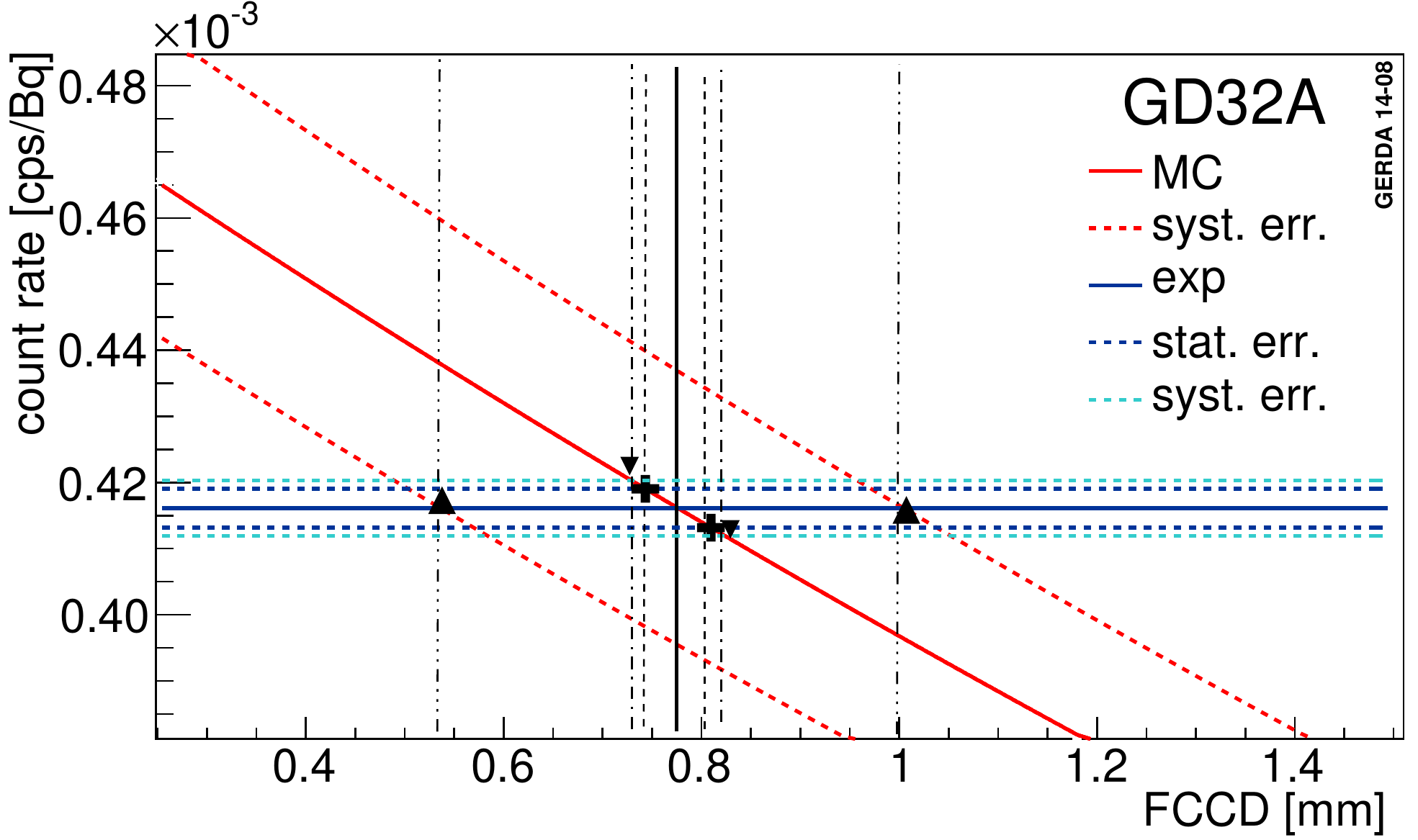}
\caption{  \label{fig:peakcountrate-method}
         Extraction of the FCCD value of detector GD32A from a comparison of
         the measured vs. simulated absolute count rate of the 1173\,keV
         $\gamma$-line of $^{60}$Co:  MC simulated value as a function of FCCD
         (red), measured observable (blue), intersection (black).
}
\end{center}
\end{figure}

\begin{figure}[t]
\begin{center}
\includegraphics[width=0.9\columnwidth]{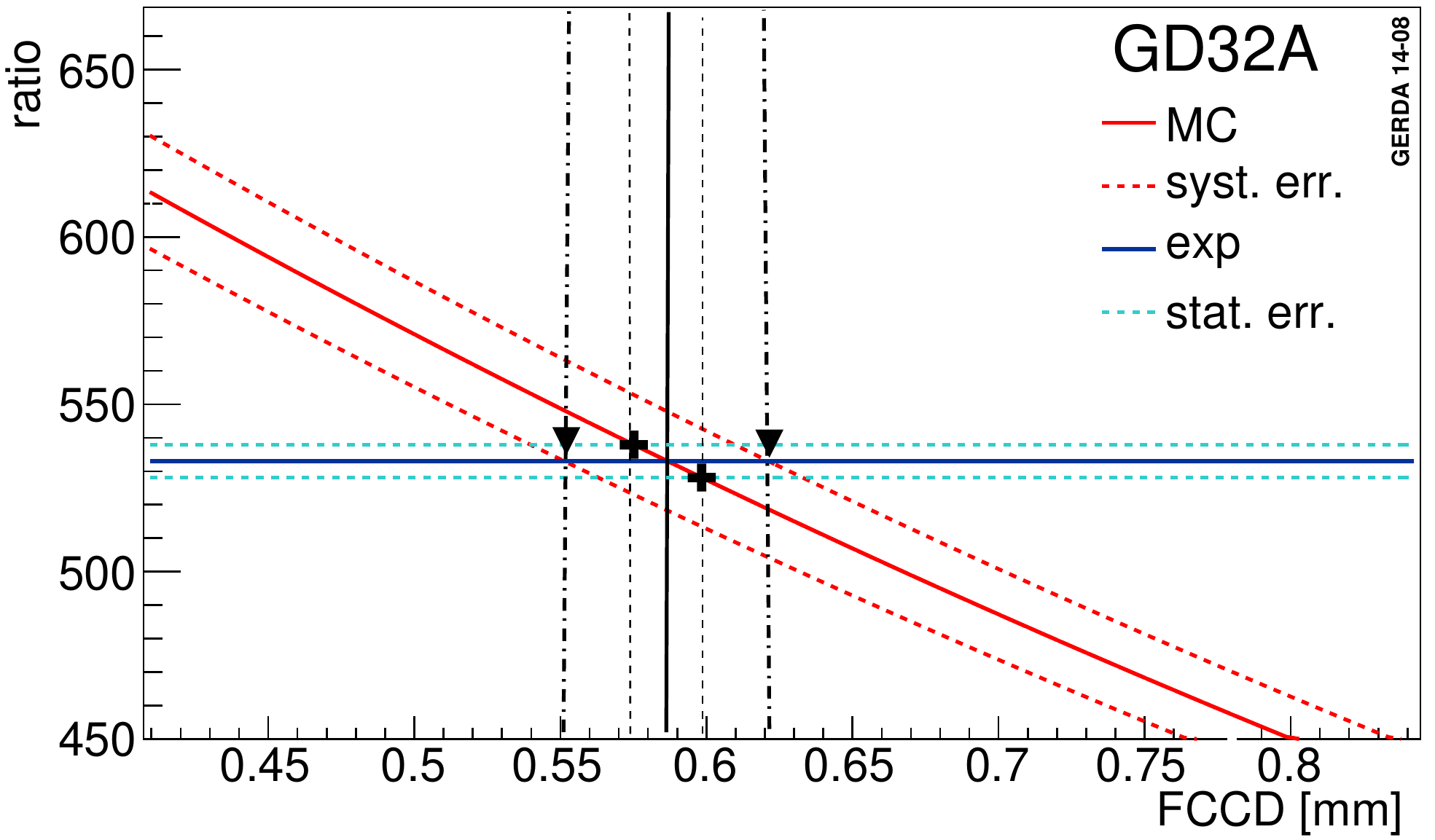}
\caption{  \label{fig:ratio-method}
         Extraction of the FCCD value of detector GD32A from a comparison of the
         measured vs. simulated ratio of two $\gamma$-line intensities for
         $^{241}$Am. Lines are color-coded as in
         Fig.~\ref{fig:peakcountrate-method}.
}
\end{center}
\end{figure}

MC simulations were performed in the \mage~\cite{bib:MAGE} software framework
based on \geant~\cite{bib:GEANT4_1,bib:GEANT4_2} version 9.4.p04. Simulations
of both source measurement types were performed for all detectors. Afterwards
the peak counts of a subset of $\gamma$-ray peaks in the experimental and MC
simulated energy spectra were evaluated by both a fitting and a counting
method. Then, two types of observables; i.e., either count rates or count rate
ratios, were extracted and plotted as a function of the FCCD. As a working
hypothesis an equal FCCD thickness on the top, lateral and bottom sides of the
detector surface was assumed. In the case of $^{60}$Co, the MC peak count
rates were plotted as a function of the FCCD. An example is given in
Fig.~\ref{fig:peakcountrate-method}. The intersection of the experimental
result and the simulated curve gives the FCCD of the detector.  This method
depends strongly on the precise knowledge of the detector dimensions, source
activity and distance of source to detector. In case of $^{241}$Am, the ratio
of the count rates in the 60\,keV peak and the summed count rates from the
neighboring peaks at 99 and 103\,keV in the measured spectrum were compared
with the corresponding ratios of MC spectra for different FCCD values. An
example is given in Fig.~\ref{fig:ratio-method}. By using the peak count
ratio, uncertainties emerging from the source-to-detector distance and the
source activity cancel out. The intersection of the measured ratio and the MC
ratios as a function of the FCCD thickness defines the average upper surface
FCCD thickness and in consequence the AV of the detector.

\begin{table*}[t]
\begin{center}
\caption{\label{tab:all_comb_only_1173_only}
       Active volume fractions ($\factvol$) of the first seven enriched BEGe
       detectors for \gerda . The systematic uncertainties are split in
       detector correlated and uncorrelated contributions. The 6th column
       presents $\factvol^\star$ where the uncertainties are added in
       quadrature. For comparison, the manufacturer specifications, which were
       provided without uncertainty, are reported in the last column.
}
\begin{tabular}{llrrrrr}
\hline
detector & $\gamma$-peak/nuclide & $\factvol$ $^{+stat +ucorr +corr}_{-stat -ucorr -corr}$ & $\factvol^\star$& Canberra\\
\hline
\noalign{\vskip 1.0mm}
GD32A & $1173$\,keV & $ 0.8962 ^{+0.0006+0.0066+0.0250}_{-0.0006-0.0060-0.0237} $ &$ 0.896 ^{+0.026}_{-0.024}$& \\
\noalign{\vskip 1.0mm}
      & $^{241}$Am & $ 0.9229 ^{+0.0015+0.0009+0.0043}_{-0.0015-0.0009-0.0041} $ & $ 0.923 ^{+0.005}_{-0.004}$ & $ 0.922 $  \\
\noalign{\vskip 1.0mm}
\hline
\noalign{\vskip 1.0mm}
GD32B & $1173$\,keV & $ 0.8887 ^{+0.0010+0.0062+0.0258}_{-0.0010-0.0056-0.0244} $ & $ 0.889 ^{+0.027}_{-0.025}$&  \\
\noalign{\vskip 1.0mm}
      & $^{241}$Am & $ 0.9059 ^{+0.0012+0.0013+0.0037}_{-0.0012-0.0013-0.0036} $ & $ 0.906 ^{+0.004}_{-0.004}$&  $ 0.899 $\\
\noalign{\vskip 1.0mm}
\hline
\noalign{\vskip 1.0mm}
GD32C & $1173$\,keV & $ 0.9069 ^{+0.0030+0.0064+0.0262}_{-0.0030-0.0058-0.0248} $ & $ 0.907 ^{+0.027}_{-0.026}$&\\
\noalign{\vskip 1.0mm}
      & $^{241}$Am &  n/a &  n/a &  $ 0.923 $ \\
\noalign{\vskip 1.0mm}
\hline
\noalign{\vskip 1.0mm}
GD32D & $1173$\,keV & $ 0.9129 ^{+0.0039+0.0062+0.0262}_{-0.0039-0.0056-0.0247} $ & $ 0.913 ^{+0.027}_{-0.026}$&  \\
\noalign{\vskip 1.0mm}
      & $^{241}$Am & $ 0.9316 ^{+0.0030+0.0006+0.0038}_{-0.0030-0.0006-0.0037} $  & $ 0.932 ^{+0.005}_{-0.005}$& $ 0.921 $\\
\noalign{\vskip 1.0mm}
\hline
\noalign{\vskip 1.0mm}
GD35A & $1173$\,keV & $ 0.9262 ^{+0.0005+0.0063+0.0261}_{-0.0005-0.0057-0.0246} $  & $ 0.926 ^{+0.027}_{-0.025}$& \\
\noalign{\vskip 1.0mm}
      & $^{241}$Am & $ 0.9369 ^{+0.0006+0.0012+0.0031}_{-0.0006-0.0012-0.0030} $   & $ 0.937 ^{+0.003}_{-0.003}$& $ 0.927 $\\
\noalign{\vskip 1.0mm}
\hline
\noalign{\vskip 1.0mm}
GD35B & $1173$\,keV & $ 0.9236 ^{+0.0013+0.0075+0.0264}_{-0.0013-0.0070-0.0249} $ & $ 0.924 ^{+0.027}_{-0.026}$&\\
\noalign{\vskip 1.0mm}
      & $^{241}$Am & $ 0.9406 ^{+0.0054+0.0016+0.0038}_{-0.0056-0.0015-0.0036} $ & $ 0.941 ^{+0.007}_{-0.007}$& $ 0.923 $ \\
\noalign{\vskip 1.0mm}
\hline
\noalign{\vskip 1.0mm}
GD35C & $1173$\,keV & $ 0.9036 ^{+0.0025+0.0079+0.0251}_{-0.0025-0.0073-0.0238} $ & $ 0.904 ^{+0.026}_{-0.025}$&\\
\noalign{\vskip 1.0mm}
      & $^{241}$Am & $ 0.9288 ^{+0.0013+0.0010+0.0042}_{-0.0014-0.0010-0.0040} $ & $ 0.929 ^{+0.005}_{-0.004}$& $ 0.926 $ \\
\noalign{\vskip 1.0mm}
\hline 
\end{tabular}
\end{center}
\end{table*}

The determined $\factvol$ values based on $^{241}$Am and the $1173\,$keV
$^{60}$Co $\gamma$-line are reported in
Table~\ref{tab:all_comb_only_1173_only}. In summary:

$(i)$ reported central values and uncertainty budgets are slightly improved
compared to a previous evaluation used in
Refs.~\cite{bib:GER13-0nbb,bib:GER13-bckg}.

$(ii)$ The manufacturer used an $^{241}$Am surface probe to estimate the
  FCCD thicknesses. Translating these results into $\factvol$ fractions, they
  are in good agreement at a 1\% level with the $^{241}$Am results of the
  current analysis.

$(iii)$ The $\factvol$ central values deduced from the $^{60}$Co
  measurements are systematically lower than the ones obtained via the
  $^{241}$Am surface tests. For two detectors the difference is at the 1\%
  level, in the other five cases it is between 2 and 3\%. On average, the
  difference is 1.9\%. 28 potential sources of systematic uncertainties were
  evaluated. The most prominent contributions are reported in
  Table~\ref{tab:systematics_both}. For each detector the total systematic
  uncertainty was divided into a detector correlated and non-correlated
  part. An example for the first category is the usage of the same calibration
  source for all detectors, which -- in case of an offset -- would cause an
  asymmetric shift in one direction for all $\factvol$ mean values. As shown
  in Table~\ref{tab:all_comb_only_1173_only}, the correlated systematic
  uncertainty of the $^{60}$Co measurements can explain the observed shift.

\subsubsection{Background rejection via pulse shape analysis}
\label{sec:vacuum_psa}

\begin{table*}[t]
\begin{center}
\caption{\label{tab:systematics_both}
         Main systematic uncertainties considered in the determination of the
         $\factvol$ fractions. All systematic contributions are given in \%
         with respect to the count rate of $^{60}$Co or to the ratio for
         $^{241}$Am except for the dead time uncertainties (*) which are first
         translated into a live time uncertainty. This depends on the
         respective live time of the single measurement, which was typically
         $>$97\%. In the case of $^{241}$Am, two sources with different
         uncertainties were used. Thus, both numbers are reported.
}
\begin{tabular}{llrr}
\hline
category & systematics & uncert. [\%] ($^{60}$Co) & uncert. [\%] ($^{241}$Am)\\
\hline
\multirow{4}{*}{MC physics processes} & \geant\ physics~\cite{bib:Geant4sys} & $\pm 4$ & $\pm 2$ \\
& $\gamma$-line intensity $1173\,$keV & $\pm 0.03$ & -\\ 
& $\gamma$-line intensity $1333\,$keV & $\pm 0.0006$ & -\\ 
& $\gamma$-line intensity $^{241}$Am & - & $\pm 1.5$ \\ 
\hline
\multirow{3}{*}{$\gamma$-ray source} & source activity & $\pm 1$  & 0\\
& source material & $\pm 0.01$ & $\pm 0/\pm 0.014$\\
& source geometry & $\pm 0.02$ & $\pm 0.013/\pm 0.016$\\
\hline
\multirow{6}{*}{detector and cryostat}& detector dimension & $\pm 2.5$ & -\\
& dist. source to endcap& $\pm 1.2$ & -\\
& endcap geometry & $\pm 0.15$ & $\pm 0.31$\\
& dist. detector to endcap& $\pm 1.0$ & -\\
& detector cup geometry & $\pm 0.06$ & $\pm 0.03$\\
& detector cup material & $\pm 0.03$ & $\pm 0.01$\\
\hline
\multirow{2}{*}{dead time} & MCA dead time (*) & $\pm 10$ & -\\
& FADC dead time (*) & $\pm 5$ & -\\ 
\hline
\multirow{1}{*}{shaping time} & shaping time & $\pm 0.2$ & -\\
\hline
\end{tabular}
\end{center}
\end{table*}

\begin{figure}[b]
\begin{center}
\includegraphics[width=0.95\columnwidth]{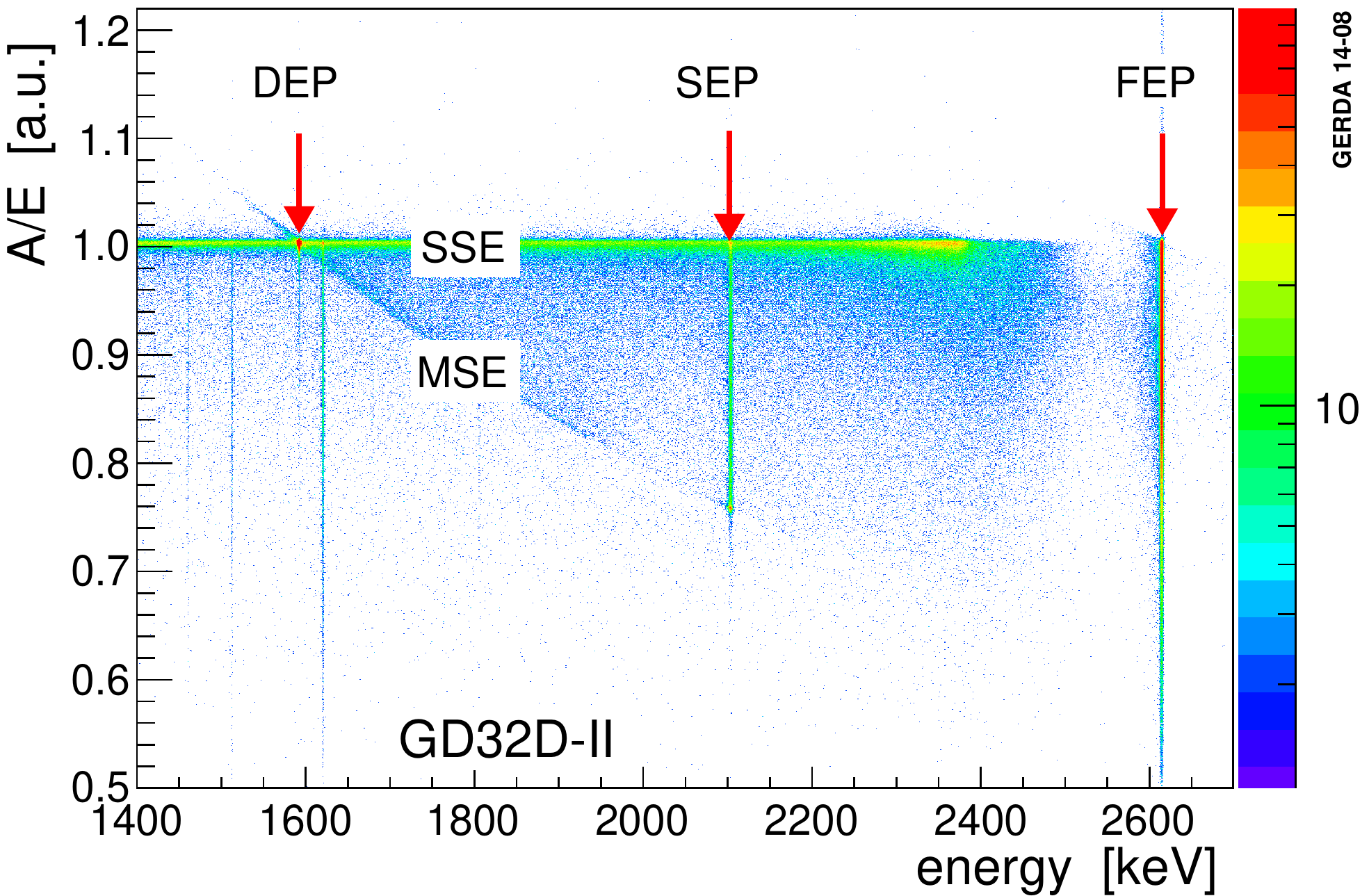}
\caption{ \label{fig:ae-vs-e_gd32dII}
           The $A/E$ ratio vs. energy $E$ from events of a $^{228}$Th
           calibration of detector GD32D-II.
}
\end{center}
\end{figure} 

For the determination of the background rejection efficiency of the new BEGe
detectors via pulse shape analysis, uncollimated $^{228}$Th calibration
sources were deployed on the outer surface of the vacuum cryostat endcaps. The
double escape peak (DEP) of the 2615\,keV photons appears at 1593\,keV. Most
DEP events are like $0\nu\beta\beta$ decays -- they are SSE unless
bremsstrahlung leads to energy losses. Contrarily, the full energy peak (FEP),
the single escape peak (SEP) and Compton continua events correspond mainly to
MSE. The obtained $A/E$ vs. $E$ plot for detector GD32D is shown exemplarily
in Fig.~\ref{fig:ae-vs-e_gd32dII}. Herein, the $A/E$ values of the events were
computed after a 10\,ns differentiation and threefold 50\,ns integration of
the charge signal.

\begin{figure}[b]
\begin{center}
\includegraphics[width=0.85\columnwidth]{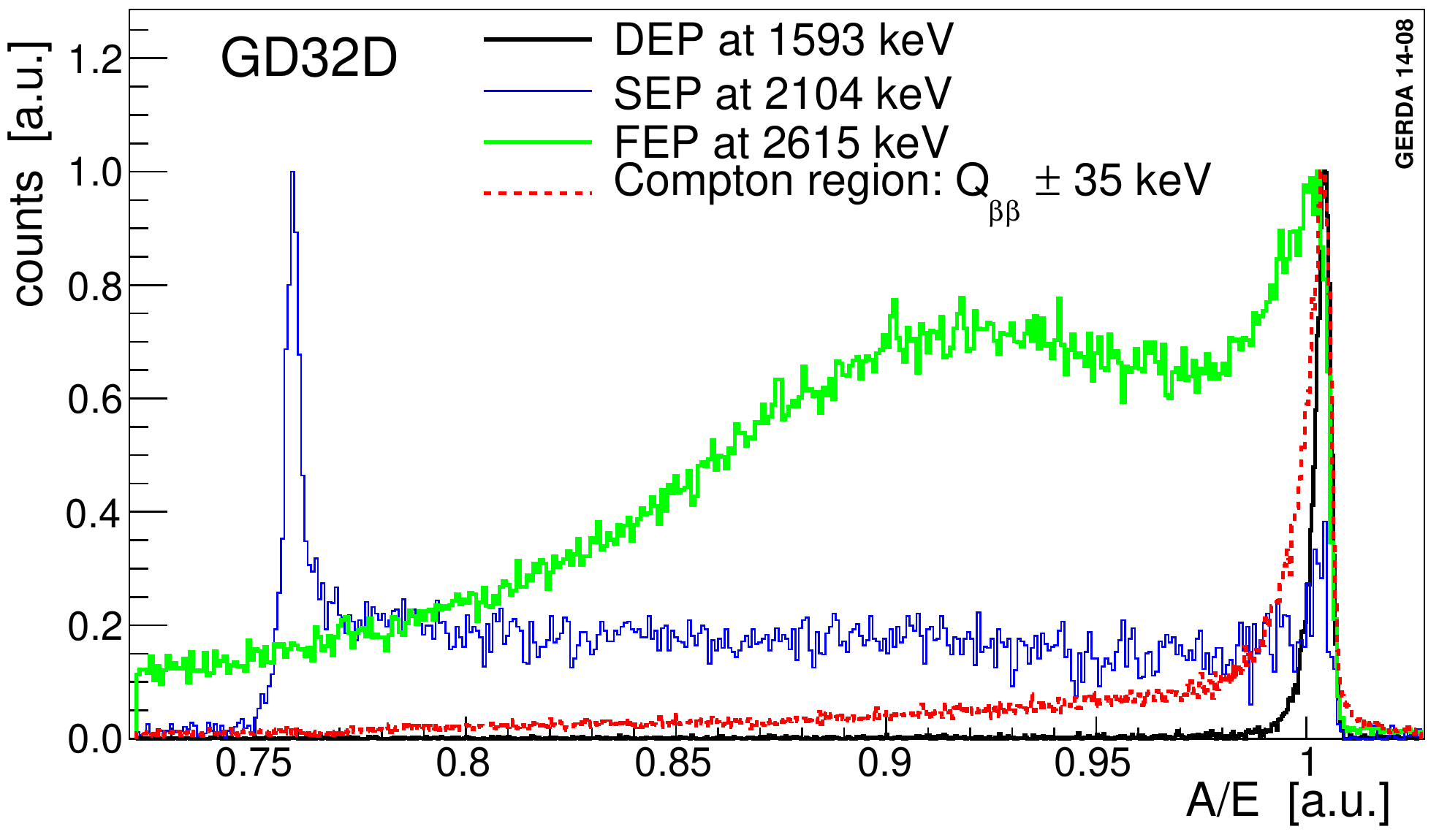}
\caption{  \label{fig:ae_single-reg_gd32dII}
             The $A/E$ distributions for selected energy
             intervals obtained from a $^{228}$Th calibration of detector
             GD32D:  full energy peak (FEP), single escape peak (SEP) and
             double escape peak (DEP) events belonging to the 2615\,keV
             $^{208}$Tl $\gamma$-line (Compton-subtracted). The $A/E$
             distribution of 
             Compton-scattered events in the 70\,keV interval around \qbb\ is
             also shown.
}
\end{center}
\end{figure} 

As a standard, a low side $A/E$ cut was set, which keeps 90\% of the events in
the Compton-background subtracted DEP. Then the survival fractions of FEP, SEP
and Compton events in the energy region between (1-2.6)\,MeV were studied (see
Fig.~\ref{fig:ae_single-reg_gd32dII}). A high side $A/E$ cut is only needed to
reject surface events occurring close to the p+ electrode.

\begin{table*}[t]
\begin{center}
\caption{ \label{tab:psa-vacuum-efficiencies}
     Gamma-ray background survival fractions (in percentages) of the first
     seven enriched BEGe detectors operated in vacuum cryostats. For
     comparison, the performance of the prototype BEGe detector denoted with
     2/B~\cite{bib:deplBEGe13} is reported as well. In the case of the
     detector GD32D, the pulse shape performance was measured before (I) and
     after (II) a rework of the groove. The $A/E$ widths $b_{A/E}$ are by trend
     correlated with the obtained PSD efficiencies. For the DEP
     the uncertainty is statistical only, while for the others the total
     uncertainty is quoted. The relative uncertainty of
     the $b_{A/E}$ fit calculation is in the range of few percent.
}
\begin{tabular}{lccccc}
\hline
detector&DEP		&SEP		&FEP		&ROI		&$b_{A/E}$\\
	&at 1593\,keV	&at 2104\,keV	&at 2615\,keV	&(2004-2074)\,keV&[\%]\\
\hline			
2/B	&90.0$\pm$0.9	&4.7$\pm$0.5	&7.0$\pm$0.4	&31.0$\pm$1.0	&0.6\\
GD32A	&90.0$\pm$0.5	&12.2$\pm$0.4	&16.3$\pm$0.7	&42.8$\pm$0.7	&1.3\\
GD32B	&90.0$\pm$0.9	&5.3$\pm$0.5	&8.8$\pm$0.5	&33.0$\pm$0.8	&0.8\\
GD32C	&90.0$\pm$1.1	&8.3$\pm$0.7	&11.3$\pm$0.4	&40.0$\pm$0.9	&1.5\\
GD32D-I	&90.0$\pm$1.1	&14.7$\pm$1.2	&22.1$\pm$1.6	&47.0$\pm$1.3	&1.5\\
GD32D-II&90.0$\pm$0.5	&5.7$\pm$0.3	&7.4$\pm$0.3	&38.3$\pm$0.8	&0.7\\
GD35A	&90.0$\pm$1.0	&7.8$\pm$0.5	&13.1$\pm$0.6	&39.5$\pm$0.7	&2.4\\
GD35B	&90.0$\pm$0.8	&5.7$\pm$0.4	&7.1$\pm$0.4	&33.0$\pm$0.8	&0.9\\
GD35C	&90.0$\pm$0.7	&10.0$\pm$0.8	&15.1$\pm$0.9	&40.3$\pm$1.2	&2.8\\
\hline 
\end{tabular}
\end{center}
\end{table*}

$A/E$ distributions of DEP events must exhibit a narrow Gaussian peak in order
to obtain a reasonable PSD efficiency. To satisfy the needs of the
\gerda\ experiment a FWHM of $\lesssim$1\% is required. However, a small tail
component of MSE with lower $A/E$ from underlying bremsstrahlung background is
allowed.  SSE populating the peak can then be disentangled from the tail
region.  The PSD results are reported in
Table~\ref{tab:psa-vacuum-efficiencies}.

\begin{figure}[b]
\begin{center}
\includegraphics[width=0.9\columnwidth]{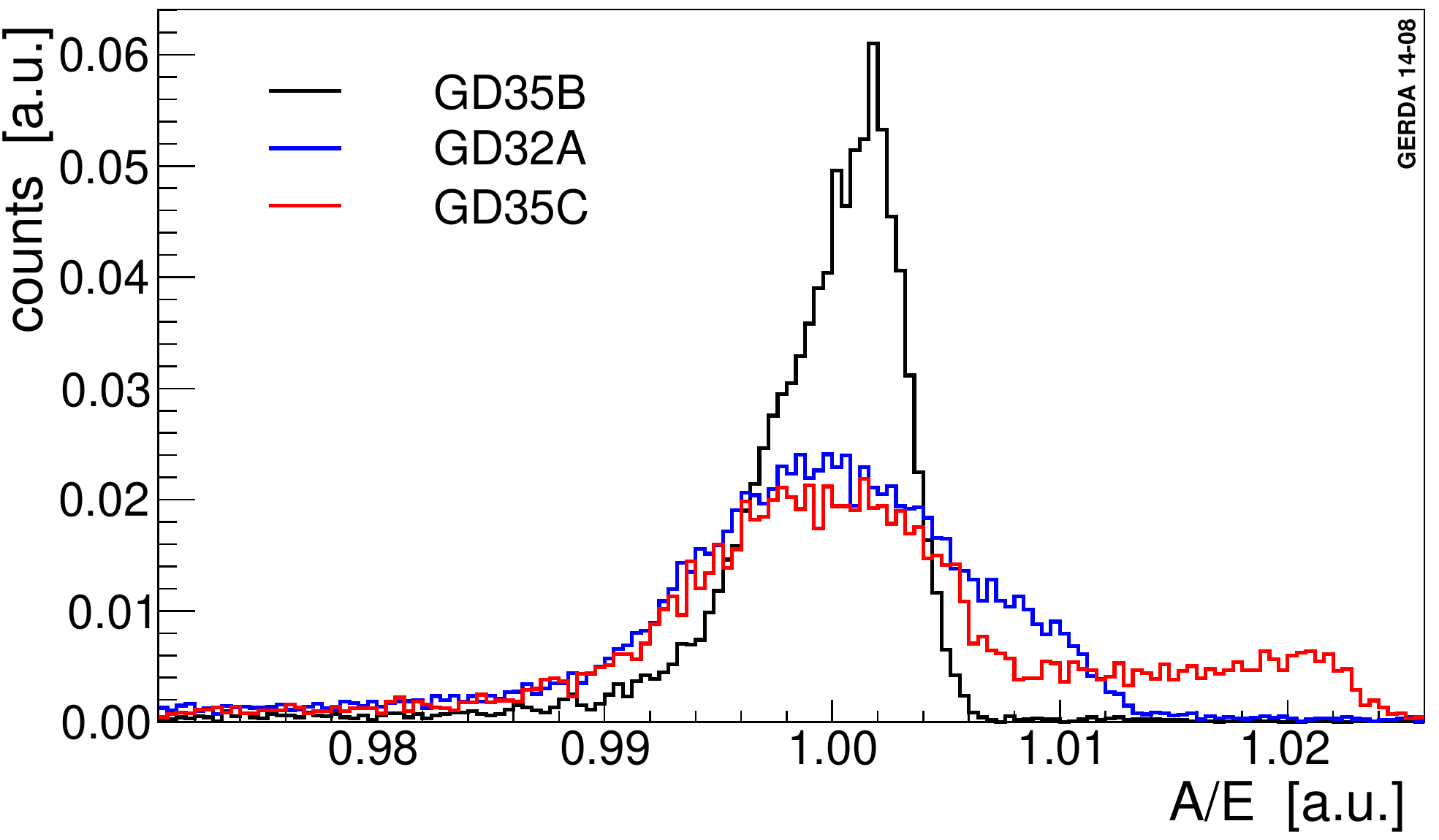}
\caption{ \label{fig:AE-resolution-of-7-BEGes}
           $A/E$ distributions of DEP events from $^{208}$Tl decays measured
           for three enriched BEGe detectors. For comparison the distributions
           were normalized. GD35B can be described by one single Gaussian with
           $\lesssim$1\% resolution plus a tail on its left side. GD32A and
           GD35C have a resolution of $\gtrsim$1\%. In addition, GD35C has a
           multiple structure.
}
\end{center}
\end{figure}

 ~\\It was found that:
$(i)$ A single $A/E$ peak was observed in GD32B, GD35B
 and in most of the prototype BEGe detectors. However, multi\-ple-peak
 structures and/or an unusual broad peak were observed in five of the enriched
 BEGe detectors (GD32(A,C,D), GD35(A,C)).
 Fig.~\ref{fig:AE-resolution-of-7-BEGes} shows one well-performing detector and
 two detectors with a deteriorated $A/E$ performance.

Setup-related artifacts were excluded by remeasuring the PSD behavior of the
prototype depleted BEGe detector denoted with 2/B. This detector was
previously measured in other laboratories and led to very similar results
(compare Table~\ref{tab:psa-vacuum-efficiencies} reported here with Table~4 in
Ref.~\cite{bib:deplBEGe13}).  Malfunctioning electronic components inside the
cryostat endcaps were excluded by reusing them for different BEGe
detectors. No notable noise effects and time instabilities were
identified. Correlations among the $A/E$ distribution shape and macroscopic
properties such as impurity concentrations or gradients, operational voltages,
crystal slice types and crystal shapes were not found. However, positive
charged compounds can be deposited in the groove on the passivation layer
between the p+ and n+ electrodes after diode production. This can deform the
electric field configuration leading to distorted drift paths. This assumption
has been substantiated by dedicated measurements with collimated $^{241}$Am
and $^{228}$Th sources. Moreover, the grooves of the detectors GD32C and GD32D
underwent chemical treatment by the manufacturer followed by heating. The
first detector was affected by operational instabilities after reprocessing,
but the second one clearly improved its PSD behavior.

The $A/E$ anomaly was further investigated with two enriched BEGe detectors
without passivation layer and operated in liquid argon, as described in
section~\ref{sec:bege_gdl-tests}.

$(ii)$ The detectors GD32B and GD35B have PSD efficiencies similar to the
prototype natural and depleted BEGe detectors~\cite{bib:deplBEGe13}. As shown
in Table~\ref{tab:psa-vacuum-efficiencies} the survival fractions of MSE lying
in the SEP and FEP are around (6-9)\%, while Compton events in the region of
interest (ROI) around \qbb\ survive at a $\sim$33\% level. The uncertainties
reported include statistical and systematic contributions. For all five other
cases the PSD efficiency is deteriorated due to the $A/E$ anomaly. In the
extreme case of GD32D, the survival probability for the $\gamma$-lines and
intervals increases to (15-22)\% and 47\%, respectively.

\subsubsection{Surface scans with $^{241}$Am sources}
\label{sec:Am-surface-scans}

Six out of the first seven \gerda\, Phase~ II BEGe detectors were scanned at
several hundred position on the top and lateral site (nomenclature referring
to Fig.~\ref{fig:bege-profile}) with a novel automatized setup consisting of a
motorized mechanical arm. The arm is equipped with a collimated 5\,MBq
$^{241}$Am source. The diameter of the collimator hole is 1\,mm. Further
details are reported in Ref.~\cite{bib:Heroica}.

The 60\,keV $\gamma$-rays emitted by the $^{241}$Am source have a typical
penetration depth of 1\,mm in germanium. This leads to an energy deposition
near the detector surface in form of a single charge cloud.

\paragraph{FCCD homogeneity and diode position:} 
For this purpose, the count rates of the 60\,keV $\gamma$-rays in single
positions along linear axes were measured.

The observed count rate drops at the edges and allowed the comparison of the
diode positions inside the cryostat endcaps with the callout in the technical
drawings of the manufacturer. No misalignment was found within $\pm$1\,mm.

\begin{figure}[b]
\begin{center}
 	\includegraphics[width=0.9\columnwidth]{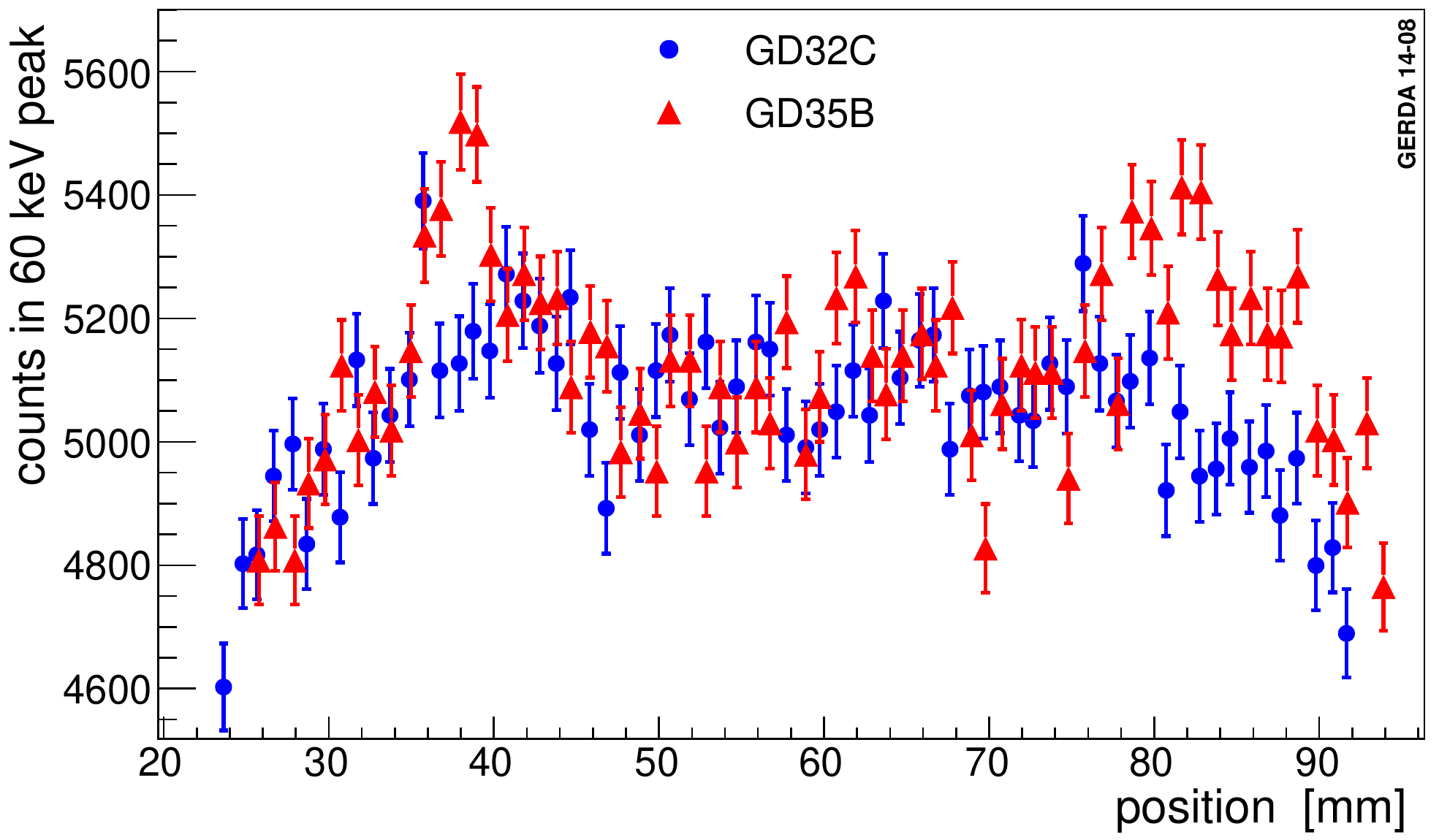}
 	\caption{ 	\label{fig:Am-scan_CCE}
             Count rate profile from a linear scan along the top side
             surface of the detectors GD32C and GD35B.
}
\end{center}
\end{figure}

Along the bulk of the diodes most of the seven detectors showed a stable count
rate.  GD32C, for instance, has an almost constant count rate profile (see
Fig.~\ref{fig:Am-scan_CCE}). A few detectors like GD35B, however, fluctuate up
to $\pm$30\% translating into a $\sim$0.1\,mm~\cite{bib:NIST} FCCD
difference. The count rate profile on the front side area of these detectors
is characterized by higher count rates at the center and at the outermost
borders. The origin of the observed fluctuation has not yet been understood.

\paragraph{Spatial dependence of the pulse shape response:}
The pulse shapes of events in a $\pm$4.5\,$\sigma$ region around the 60\,keV
$\gamma$-peak were investigated for every scanned point. Especially the
variation of the $A/E$ peak positions over the entire surface was
studied. Detectors not affected by the $A/E$ anomaly; i.e., the detectors
GD32B and GD35B were measured to have a rather good $A/E$ peak stability with
variations of 1\%. The other enriched BEGe detectors showed stronger
deviations up to 4.5\% going from the exterior towards the center. This
observation is consistent with the working hypothesis that the origin of the
$A/E$ anomaly is due to non homogeneously distributed charge carriers on the
passivation layer (see section~\ref{sec:vacuum_psa}).

\begin{figure}[t]
\begin{center}
    \includegraphics[width=0.9\columnwidth]{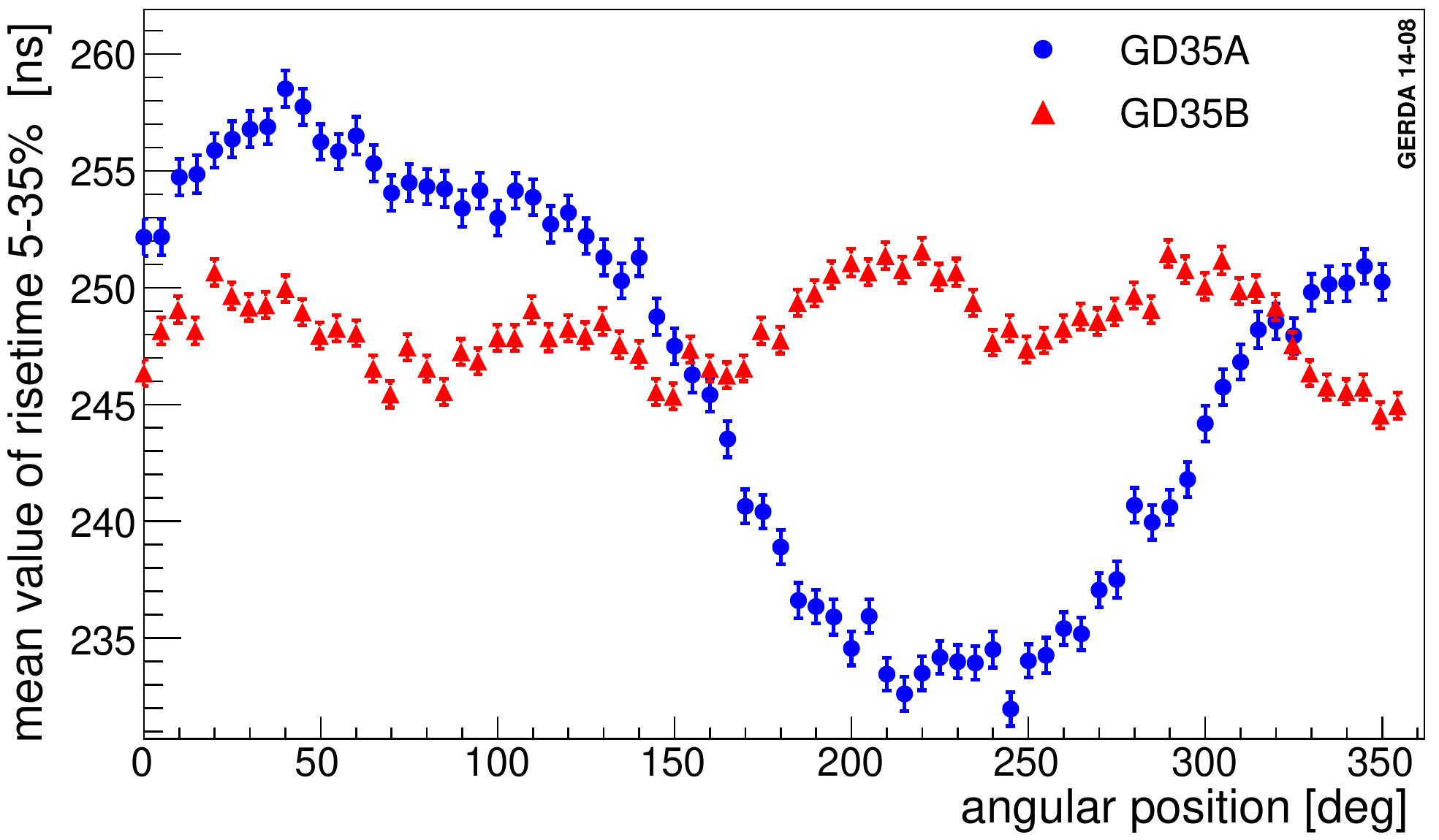}
    \caption{    \label{fig:Am-scan_AE-RT-oscillation}
          Mean value of the (5-35)\% interval of the rise time slope of the
          pulses registered along a lateral side scan of GD35A and GD35B.
}
\end{center}    
\end{figure}

A similar behavior was observed for the mean (5-35)\% rise time interval of
the registered pulses, which represents the main drift path through the
crystal from the interaction point to the p+ electrode. Detectors GD35B (see
Fig.~\ref{fig:Am-scan_AE-RT-oscillation}), GD32B and 2/B indicated a
90$^\circ$ oscillation due to the different drift mobilities for holes along
the axes of the faced-centered cubic crystal lattice of
germanium~\cite{bib:Xtal-structure}.  However, for the other enriched BEGe
detectors such as GD35A (Fig.~\ref{fig:Am-scan_AE-RT-oscillation}) the
90$^\circ$ oscillation was overwhelmed by a much larger 180$^\circ$
oscillation most likely caused by a one-sided concentration of charges in the
groove.

\section{Detector performance in liquid argon}
\label{sec:performance_lar}

Since the beginning, \gerda\ has foreseen the parallel usage of different Ge
detector designs.  The long-term stability and the pulse shape performance of
BEGe detectors have been studied with prototype natural and depleted BEGe
detectors for vacuum cryostat
operation~\cite{bib:deplBEGe13,bib:bege_pulse_shape} and in
LAr~\cite{bib:BEGe-LAr_2010}. It was also desirable to confirm the low
intrinsic background of the new detectors -- in particular for possible
surface $\alpha$-contamination potentially introduced during manufacture. For
this check an ultra-low background environment, such as inside the
\gerda\ set-up, was necessary.  Therefore, the \gerda\ collaboration decided
to operate five of the new enriched BEGe detectors already during Phase~I of
the experiment. This also allowed the study of their operational stability
over a period of $\sim$320\,d as well as their PSD {\it in situ}. The results
are presented in section~\ref{sec:bege-gerda-phaseI}. The positive performance
permitted to add 3~kg to the total mass in \gerda\ Phase~I.

The two remaining enriched BEGe detectors which were affected by the $A/E$
anomaly (section~\ref{sec:vacuum_psa}) under vacuum conditions were reworked
in order to investigate possibilities to improve the PSD performance. For this
purpose, the passivation layer of these detectors was removed. The results are
discussed in section~\ref{sec:bege_gdl-tests}.

\subsection{Operation of enriched BEGe detectors in \gerda\ Phase~I}
\label{sec:bege-gerda-phaseI}

\begin{figure}[t]
\begin{center}
\includegraphics[scale=0.9]{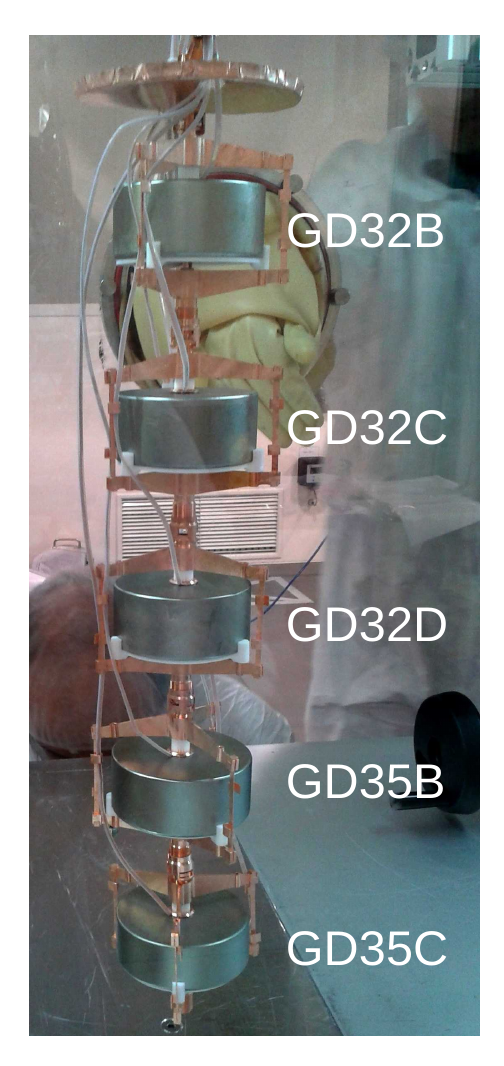}
\caption{\label{fig:bege-string} 	
       The string of five enriched BEGe detectors before deployment
       into \gerda\ (July 6, 2012).
}
\end{center}
\end{figure}

Detectors GD32B, GD32C, GD32D, GD35B and GD35C, were mounted into the
\gerda\,cryostat without prior removal of the passivation layer surrounding
the read-out electrode. Their configuration is shown in
Fig.~\ref{fig:bege-string} prior to their insertion into the \gerda\ cryostat.
Data taking relevant for the Phase~I analysis started on July~8, 2012, and
stopped on May 21, 2013.

\begin{figure*}[t]
\begin{center}
\includegraphics[width=.64\columnwidth]{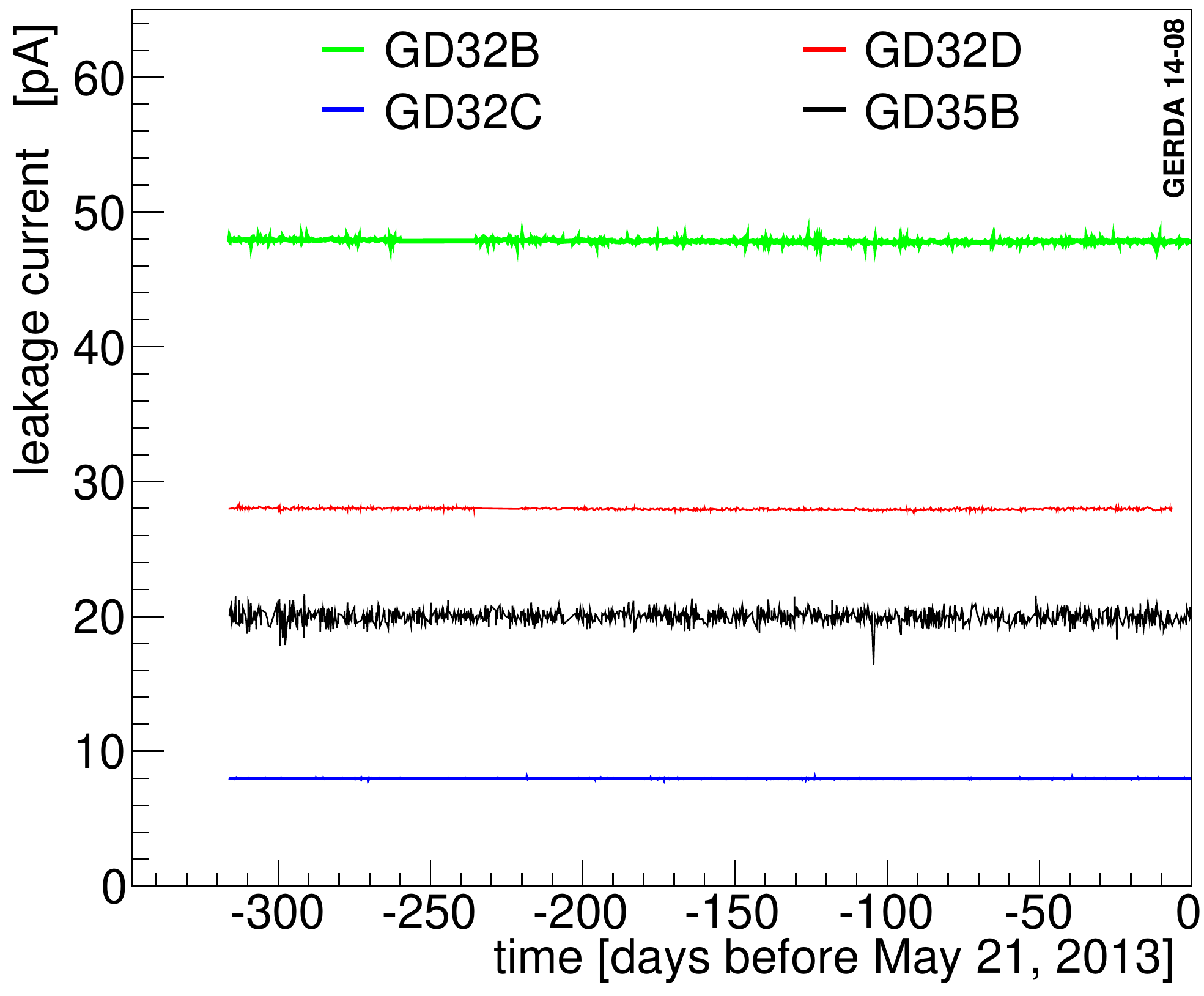}
\hfill
\includegraphics[width=.64\columnwidth]{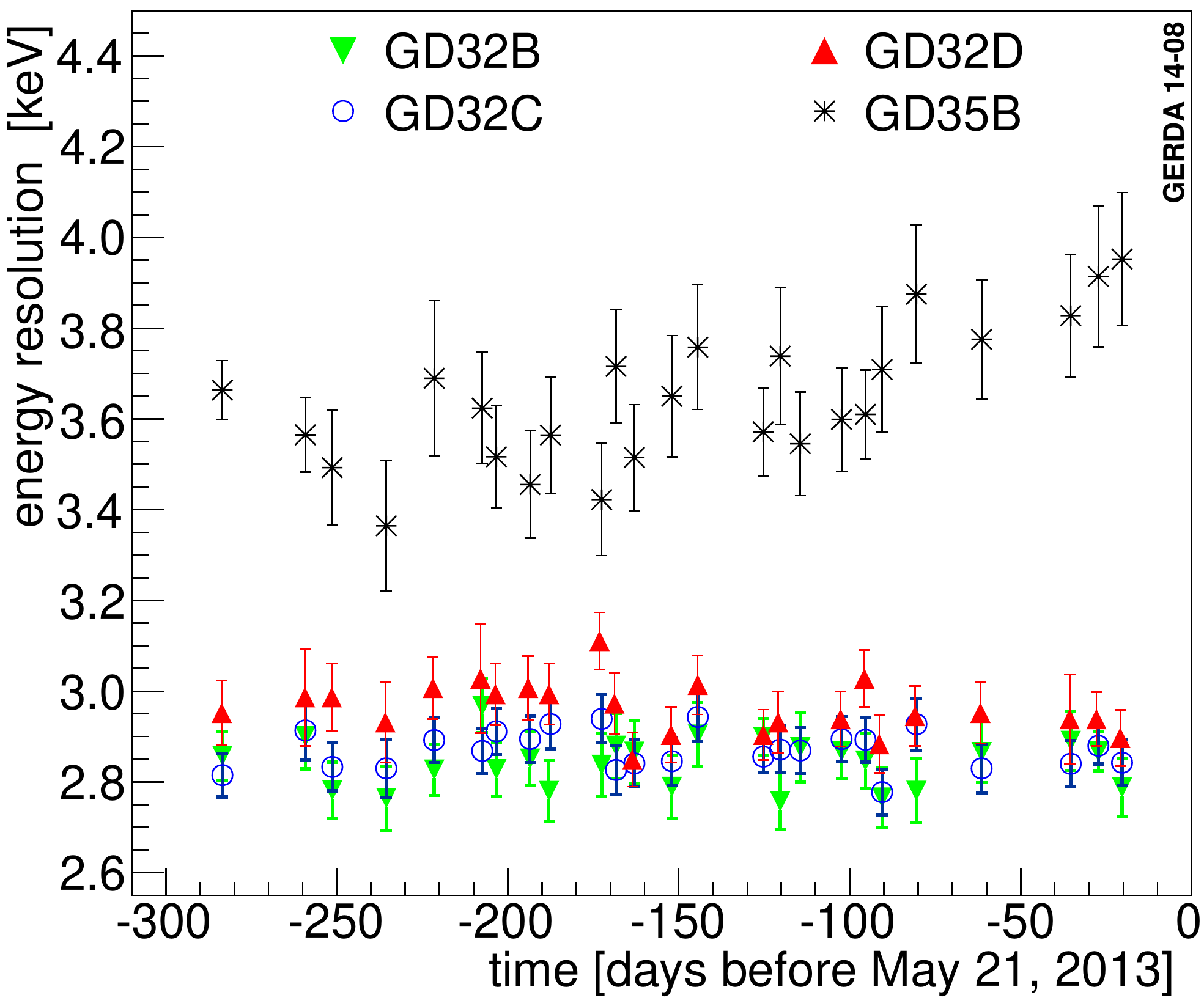}
\hfill
\includegraphics[width=.62\columnwidth]{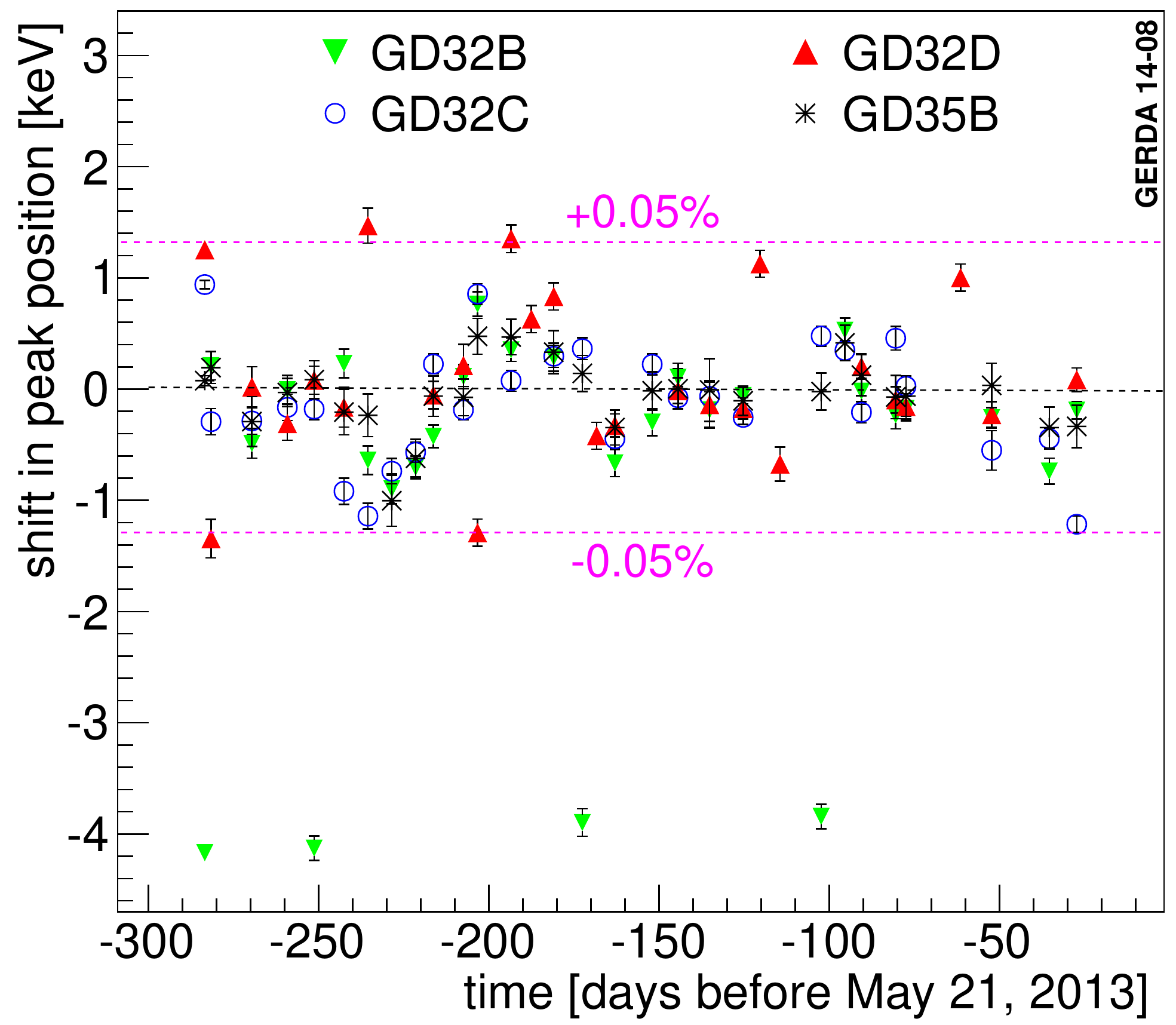}
\caption{  \label{fig:performance_bege_phaseII}
          Operational stability of the four enriched BEGe detectors operated
          in LAr during \gerda\ Phase~I: (left) Stability of leakage current
          $I_{l}$. The systematic uncertainty of $\pm$4\,pA is not
          shown. (middle) Time dependency of the energy resolution $\Delta E$
          expressed in terms of FWHM of the $^{208}$Tl $\gamma$-line at
          2615\,keV. (right) Peak position variability of the $^{208}$Tl
          $\gamma$-line at 2615\,keV.
}
\end{center}
\end{figure*} 

All five enriched BEGe detectors were operated at the same voltage of
3500\,V. According to Table~\ref{tab:ene-res-depl-volt} this voltage is equal
or even larger than the voltages recommended by the manufacturer. According to
the \gerda\ characterization measurements in section~\ref{sec:hv_scans} this
voltage guarantees a maximum active volume and an optimal energy resolution
for all detectors. Two detectors, GD35B and GD35C, were connected to the same
HV line. Both detectors showed problems during operation: GD35B had
microphonic noise leading to poor energy resolution, while detector GD35C
experienced larger gain instabilities, which prevented useful data
collection. This was possibly caused by an improper contact of the signal
cable to the p+ electrode. Ignoring GD35C, the performance of the remaining
four enriched BEGe detectors operated in \gerda\ Phase~I is summarized in the
following sections.

\subsubsection{Stability of operational and spectroscopic parameters}
\label{sec:bege-gerda_ene-res_peak-stab_leak-cur} 

\paragraph{Leakage current:}
The leakage current $I_{l}$ was calculated from the voltage drop across the
feedback resistor. It was monitored continuously for the four operational
enriched BEGe detectors. Initially, $I_{l}$ was measured via the test point
voltage of the detectors. During the long-term operation $I_{l}$ was extracted
from the baseline of recorded events. As demonstrated in
Fig.~\ref{fig:performance_bege_phaseII}(left), $I_{l}$ of all detectors was
acceptably low. The systematic uncertainty of $\pm$4\,pA is not
shown. Additionally, $I_{l}$ was almost constant during the entire data
collection time.

Almost weekly $^{228}$Th source calibration tests did not increase $I_{l}$. In
the past, source measurements of several days duration conducted on prototype
semi-coaxial detectors with a passivation layer led to a significant increase
of $I_{l}$~\cite{bib:Marik2008}. For the passivated BEGe detectors in
\gerda\ Phase~I this was probably avoided by collecting enough statistics
within only few hours.

\paragraph{Energy resolution:}
The energy resolution $\Delta E$ was determined by irradiating the four
enriched BEGe detectors with $^{228}$Th sources in LAr, and by recording and
processing data according to Ref.~\cite{bib:GER13}. In total, 25 different
calibration data sets for all detectors were available. The corresponding FWHM
of the $^{208}$Tl $\gamma$-line at 2615\,keV are depicted in
Fig.~\ref{fig:performance_bege_phaseII}(middle). All detectors had a good and
stable $\Delta E$ over the entire period except GD35B. The averaged $\Delta E$
of the other three detectors ranged between 2.8 and 3.0\,keV. The averaged
$\Delta E$ of GD35B was at 3.6\,keV, and there was an increase of 0.4 keV over
this time period.

The $\Delta E$ values of the four enriched BEGe detectors operated in LAr were
30\% worse than the values obtained for the same detectors operated in vacuum
cryostats (see Table~\ref{tab:ene-res-depl-volt}). This was expected, since
the signal cable length in LAr between the read-out electrode and the first
stage of the preamplifier, the field-effect transistor (FET), was longer by
$\sim$30\,cm. For \gerda\ Phase~II, however, it is planned to place the FET
closer to the read-out electrode. In addition, new offline energy
reconstruction algorithms are under development that should further improve
the energy resolution.

Finally, compared to the semi-coaxial Ge detectors operated in LAr with almost
identical read-out electronics, the four BEGe detectors had a $\sim$30\%
better energy resolution~\cite{bib:GER13-bckg}.

\paragraph{Peak position stability:} The energy scale stability of the four
enriched BEGe detectors was measured by means of the $^{208}$Tl $\gamma$-peak
at 2615\,keV. As shown in Fig.~\ref{fig:performance_bege_phaseII}(right) the
peak position was relatively stable with deviations mostly within
$\pm$0.05\%. Temporary instabilities in the energy scale of individual
detectors were identified by analyzing the response to regularly injected
charge pulses into the input of the amplifiers. Possibly, the signal and/or
the HV contact was not stable. Data affected by the instabilities were
excluded in the data analysis.

\begin{table*}[t]
\begin{center}
\caption{\label{tab:psa-LAr-efficiencies}
     Gamma-ray background survival fraction (in percentages) of the four
     enriched BEGe detectors operated in LAr during
     \gerda\ Phase~I. The relative uncertainty from the $A/E$ width
     $b_{A/E}$ fit calculation is in the range of few percent.  For the DEP
     the uncertainty is statistical only, while for the others the total
     uncertainty is quoted. For a direct comparison of the FEP and SEP
     efficiencies with Table~\ref{tab:psa-vacuum-efficiencies} one should
     consider a systematic contribution of 1\% due to geometric effects from
     different detector-source configurations in LAr and vacuum.
}
\begin{tabular}{lccccc}
\hline
detector&DEP		&SEP		&FEP		&ROI 	& $b_{A/E}$\\ 
	&at 1593\,keV &at 2104\,keV &at 2615\,keV &(2004-2074)\,keV &[\%]\\
\hline
GD32B   &90.0$\pm$0.9 &11.4$\pm$0.7 &15.1$\pm$1.0 &44.3$\pm$1.0 &1.5 \\ 
GD32C   &90.0$\pm$0.8 &11.3$\pm$0.7 &14.7$\pm$0.9 &45.9$\pm$1.0 &1.7 \\
GD32D   &90.0$\pm$1.1 &10.2$\pm$0.7 &14.2$\pm$0.9 &45.2$\pm$1.2 &1.6 \\ 
GD35B   &90.0$\pm$1.5 &9.9$\pm$1.3  &16.2$\pm$1.5 &46.4$\pm$2.0 &1.9 \\
\hline 
\end{tabular}
\end{center}
\end{table*}

\subsubsection{Pulse shape performance}
\label{sec:bege-gerda_psa}

The pulse shape behavior of the four enriched BEGe detectors has already been
discussed in the context of the pulse shape methods developed for
\gerda\ Phase~I data analysis~\cite{bib:GER13-psd}. It has been pointed out,
that the mean value $\mu_{A/E}$ of the Gaussian components describing the
$A/E$ distributions was affected by two time variations. Firstly, an
exponentially decreasing $\mu_{A/E}$ with a time period of $\sim$1 month was
observed. The size of the total drift depended on the detector and varied from
1 to 5\% (largest for GD32B). Secondly, a $\mu_{A/E}$ shift of 1\% to higher
values during $^{228}$Th source calibrations was observed. These instabilities
were quantified and time-corrections were applied to the $A/E$ distributions
of calibration and physics data. The origin of the dynamic processes is still
under investigation. The observed dynamic $A/E$ drift might originate from
charges present on the passivated groove, which neutralize or dissolve into
the LAr over several months of operation.

The width $b_{A/E}$ of the $A/E$ distributions obtained from all calibration
data turned out to be $\gtrsim$1.5\% for the four BEGe detectors. This was
sufficient for Phase~I, however, a value $\le$1\% is aimed for Phase~II.  It
was demonstrated that a $b_{A/E}$ of 1\% can be achieved with BEGes in LAr
(section~\ref{sec:bege_gdl-tests}).

\begin{figure}[b]
\centering
  \includegraphics[width=0.9\columnwidth]{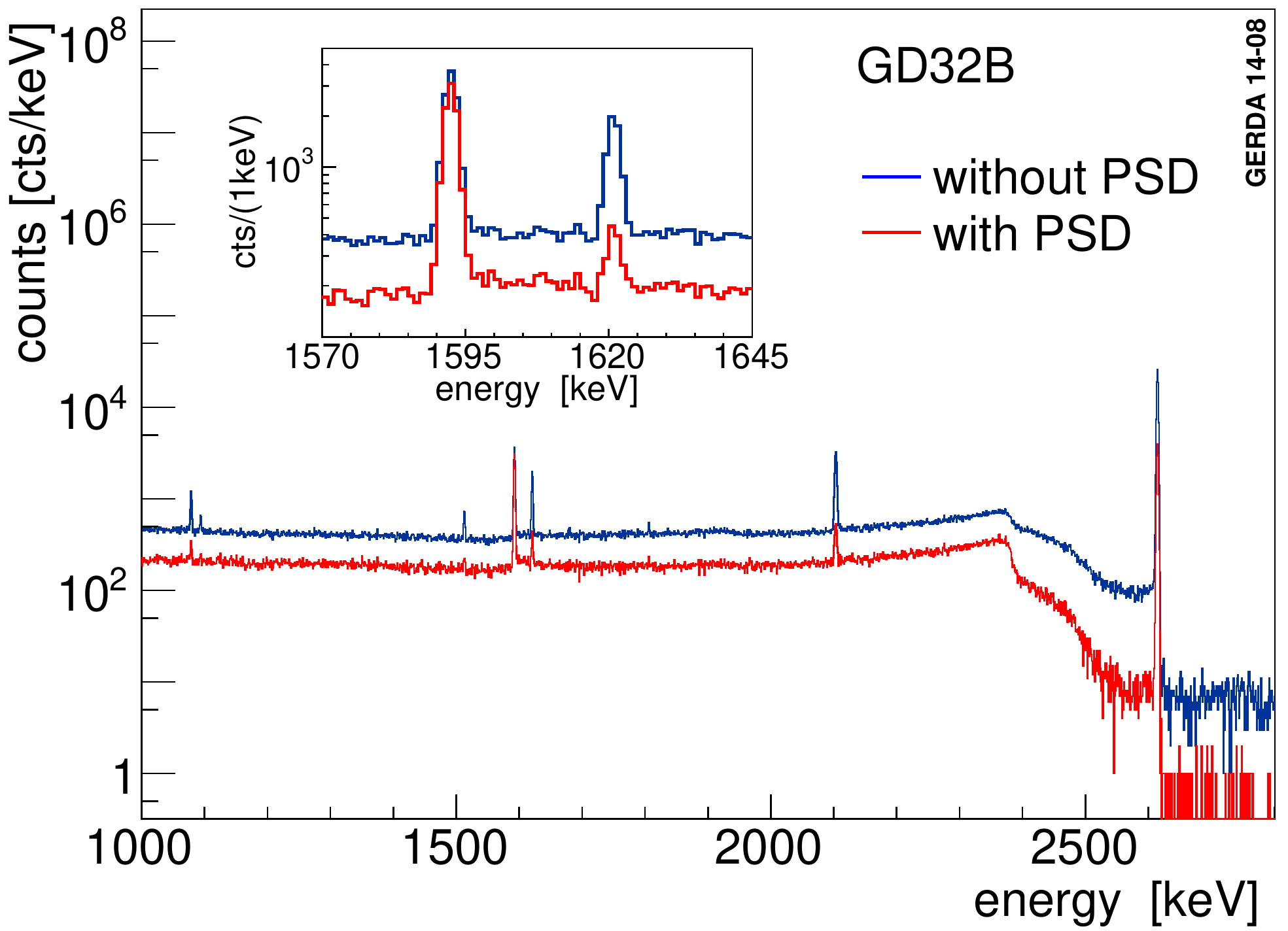}
  \caption{\label{fig:Gerda_BEGe_psa-gamma-rejection}
          Application of the $A/E$ pulse shape method on \gerda\ $^{228}$Th
          calibration data collected with detector GD32B. SSE in the DEP at
          1593\,keV were retained at a fixed 90\% level. The resulting
          survival fraction of other $\gamma$-lines and Compton regions are
          reported in Table~\ref{tab:psa-LAr-efficiencies}.
}
  \end{figure}

In contrast to the Phase~I analysis the acceptance of SSE-like DEP events at
1593\,keV was fixed at 90\%, in order to compare the performance of different
detectors and set-ups, in particular with the vacuum tests. The resulting
$\gamma$-ray background survival fractions are reported in
Table~\ref{tab:psa-LAr-efficiencies}. The energy spectrum before and after the
application of the PSD cut is shown exemplarily for detector GD32B in
Fig.~\ref{fig:Gerda_BEGe_psa-gamma-rejection}. After Compton subtraction of
the SEP and the FEP peaks, 89\% and 85\% of the $\gamma$-ray events are
rejected in LAr. The values obtained from the same detectors operated in
vacuum (see Table~\ref{tab:psa-vacuum-efficiencies}) are slightly higher, even
considering an additional $\sim$1\% uncertainty due to different
source-detector distances in the two configurations. This was expected due to
the less favorable electronics setup in \gerda\ Phase~I compared to the
operation in vacuum cryostats. Concerning the Compton-scattered $\gamma$-rays
with energies around \qbb\, the events are rejected at $\sim$55\% in
LAr. Since the Compton continuum is dependent on the source distance and
position, this value cannot be directly compared with the vacuum test results.

Finally, the PSD method was applied to the physics data of \gerda\ Phase~I
~\cite{bib:GER13-0nbb}. Herein, the $A/E$ acceptance cut on SSE-like events at
\qbb\ was optimized differently than the one used for the $^{228}$Th
calibration data collected in vacuum and in LAr.  Firstly, events below and
above the SSE band were cut. Secondly, the cut was relaxed to keep more SSE-like
events. Thirdly, the cut was at a fixed normalized $A/E$ value instead of a
fixed DEP acceptance; i.e. 0.965$<A/E<$1.07. The final $A/E$ rejection
efficiency of all background events in the ROI around \qbb\ was $\sim$82\%,
while (92$\pm$2)\% of \onbb\ events would survive the $A/E$
cut~\cite{bib:GER13-0nbb,bib:GER13-psd}.

\subsubsection{Background examination}
\label{sec:detector-contamination} 

The background sources affecting the enriched BEGe detectors in \gerda\ Phase~I
have already been discussed in Ref.~\cite{bib:GER13-bckg}.

Gamma-rays induced by $^{214}$Bi and $^{228}$Th decays occuring in assembly
materials close to the detectors were found to be the prevailing contaminants.

The predicted background induced by detector-in\-trin\-sic decays of the two
dominant cosmogenic radioisotopes, $^{68}$Ge and $^{60}$Co, was presented in
section~\ref{sec:intro-cosmogenics}. The spectral fit of the \gerda\ Phase~I
data revealed a good agreement with the expectations. The outcome reinforced
the confidence that -- with the knowledge of the exposure history of a
detector to cosmic radiation -- this background contribution can be reliably
controlled.

The following paragraphs review some aspects of two other important
contaminants which are not induced by gamma-rays and therefore their
suppression relies on detector PSD instead of LAr scintillation
anti-coincidence.

\paragraph{Cosmogenic $^{42}$Ar in LAr:}
The most critical external background for BEGe detectors in \gerda\ originates
from $\beta^-$-emission of the $^{42}$Ar daughter nuclide $^{42}$K. The
$\beta$-particles with energies up to 3525\,keV have a total stopping power of
up to 1.5\,MeV$\cdot$cm$^2$/g in natural germanium ~\cite{bib:NISTbeta} and
depending on distance are able to penetrate the n+ FCCD.  According to
Ref.~\cite{bib:GER13-bckg}, $^{42}$K $\beta$-events are expected to contribute
at a $\sim$55\% level to the total $BI$ of
(4.2$\pm$0.7)$\times$10$^{-2}$\,\ctsper\, of the four BEGe detectors in
\gerda\ Phase~I~\cite{bib:GER13-psd}. Thus, the $^{42}$Ar concentration should
be either reduced or the induced background signals suppressed by PSD
techniques.

Since the $^{42}$K $\beta$-surface events have typically long rise times it is
possible to suppress them efficiently via PSD (see also Fig.~12 in
Ref.~\cite{bib:GER13-psd}). The $A/E$ method led to a $^{42}$K survival
fraction of few per cent, as expected from experimental investigations in
Ref.~\cite{bib:2012lazzaro}. The total $BI$ of the BEGe detectors operated in
\gerda\ Phase~I was reduced to (5$^{+4}_{-3}$)$\times$10$^{-3}$\,\ctsper\,by
the PSD cut~\cite{bib:GER13-0nbb}. In Phase~II the BI can be further reduced
using a transparent nylon cylinder around the detector strings.

\paragraph{Surface $\alpha$ contamination:}
A second detector-intrinsic background is given by a potential
$\alpha$-contamination of $^{226}$Ra daughter nuclides on the detector
surface. Alpha-particles have a short range in germanium of the order of tens
of \mum, but are able to penetrate the 140\,nm passivation layer in the
detector groove and the 400\,nm thick p+ electrode, respectively.

During the detector production at Canberra no special precautions were adopted
such as handling under radon-free atmosphere. The four enriched BEGe detectors
operated in the low background environment of the \gerda\ cryostat allowed to
quantify the grade of the surface $\alpha$-contamination. In the data set used
for \gerda\ Phase~I data analysis~\cite{bib:GER13-0nbb} approximately 40
$\alpha$-like events with energies above 3\,MeV were collected with the
enriched BEGe detectors.

\begin{figure}[b]
\centering
  \includegraphics[width=0.95\columnwidth]{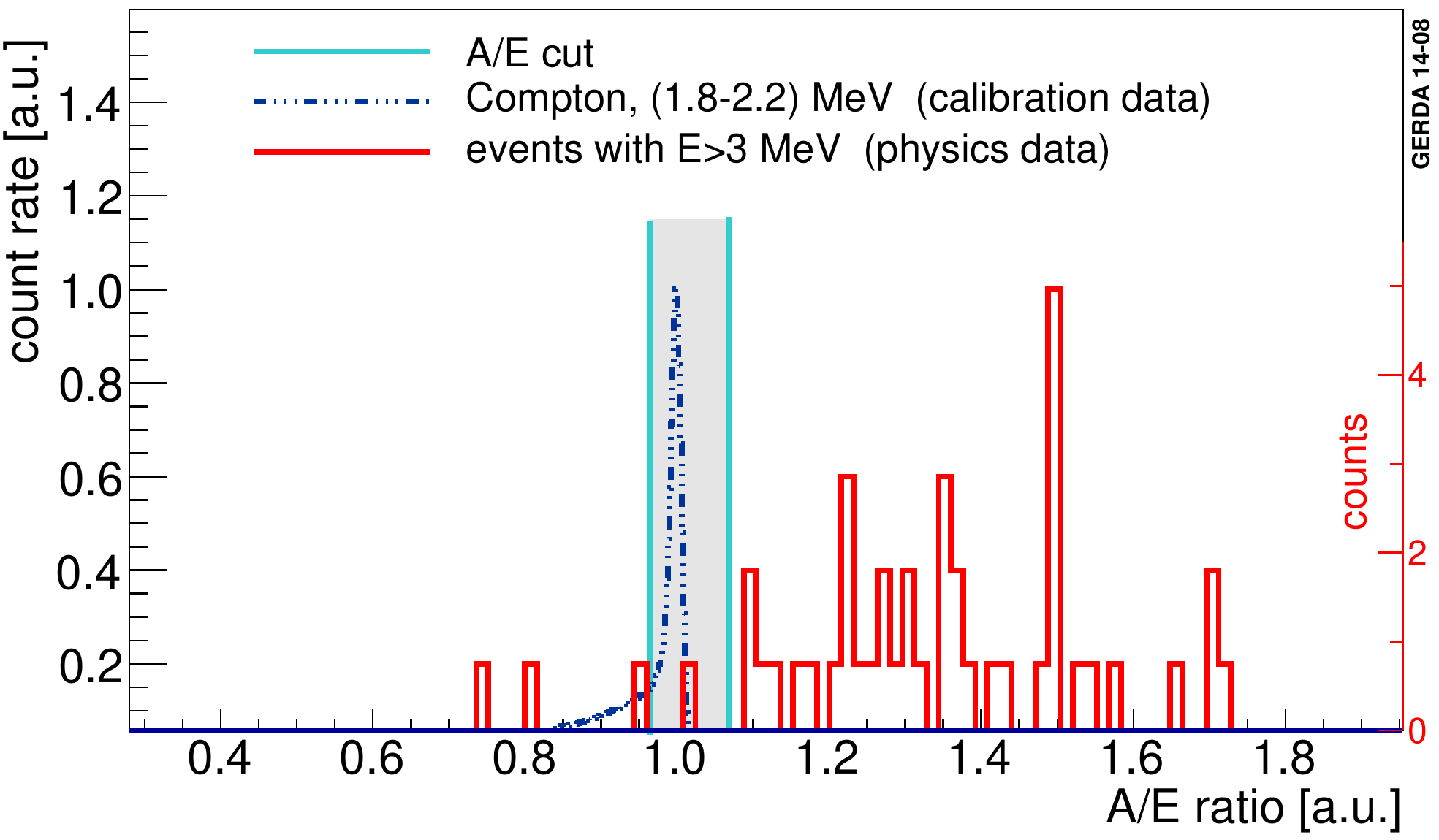}
  \caption{\label{fig:BEGe_alpha-contamination}
         $A/E$ representation of the 43 events with energies above 3.0\,MeV,
         which were registered with the four enriched BEGe detectors during
         \gerda\ Phase~I. Events inside the gray area lie within the SSE band
         and are accepted by the PSD cut.
}
\end{figure}

Figure~\ref{fig:BEGe_alpha-contamination} depicts the events from Fig.~10 in
Ref.~\cite{bib:GER13-psd}, which have energies above 3\,MeV and were projected
on the $A/E$ axis. Out of 43 candidates 39 were found with high $A/E$ values
well beyond the threshold for \onbb-like SSE. These are mostly due to
$\alpha$-events in agreement with the $A/E$ expectation from an
$\alpha$-source irradiation of a p+ electrode of a BEGe detector inside a
custom-made vacuum cryostat~\cite{thesis_matteo}. Out of these, 14 events with
$A/E$$>$1 show strong pairwise time correlations. Since each pair or triple
occurs in the same detector with a very similar $A/E$, it is suggestive that
consecutive decays in $\alpha$ chains were observed which has caused some of
the spikes in Fig.~\ref{fig:BEGe_alpha-contamination}. A small fraction might
also be due to $^{42}$K $\beta$-surface events. Moreover, one candidate
populating the SSE band has a SSE like pulse shape, while three other events
with $A/E$ $<$0.965 are identified in one case as a MSE and in two cases as
surface background events with long rise times.  Despite these few
contaminating events from non-alpha sources,
Fig.~\ref{fig:BEGe_alpha-contamination} clearly demonstrates that surface
$\alpha$ events are efficiently rejected by the A/E cut.
\subsection{Performance of BEGe detectors without passivation layer}
\label{sec:bege_gdl-tests}

As discussed in sections~\ref{sec:vacuum_psa} and~\ref{sec:bege-gerda_psa},
the observed pulse shape degradation can be caused by charge carriers present
or collected over time on the surface of the passivation layer between the
anode and cathode of the diodes.

To examine if the effect is related to environmental conditions and can be
improved for the operation in LAr, the detectors GD32A and GD35A were tested
in the \gerda\ Detector Laboratory (GDL)~\cite{bib:Marik-thesis} at LNGS.
Both detectors were deployed in the same LAr test cryostat that has already
been used for a long-term test with a prototype depleted BEGe detector which
was not passivated~\cite{bib:BEGe-LAr_2010}. The two enriched BEGe detectors
were operated in the standard configuration used in \gerda; i.e., bare
detectors in LAr without passivation on the groove surface.  According to
Ref.~\cite{bib:Marik-thesis,bib:Marik2008} this is necessary in order to
guarantee stable operation in LAr and a negligible increase of the leakage
current.

The GDL LAr tests of the two detectors revealed a slightly increased energy
resolution compared to the vacuum tests: in the case of detector GD32A it
increased from 2.5 to 2.8\,keV at 2615\,keV, while in the case of GD35A it
deteriorated from 2.4 to 3.0\,keV. This was expected, since the distance of
the first amplifying signal stage to the read-out electrodes was $\sim$60\,cm
in LAr and thus larger than in the vacuum setup.

\begin{figure}[t]
\centering
  \includegraphics[width=0.95\columnwidth]{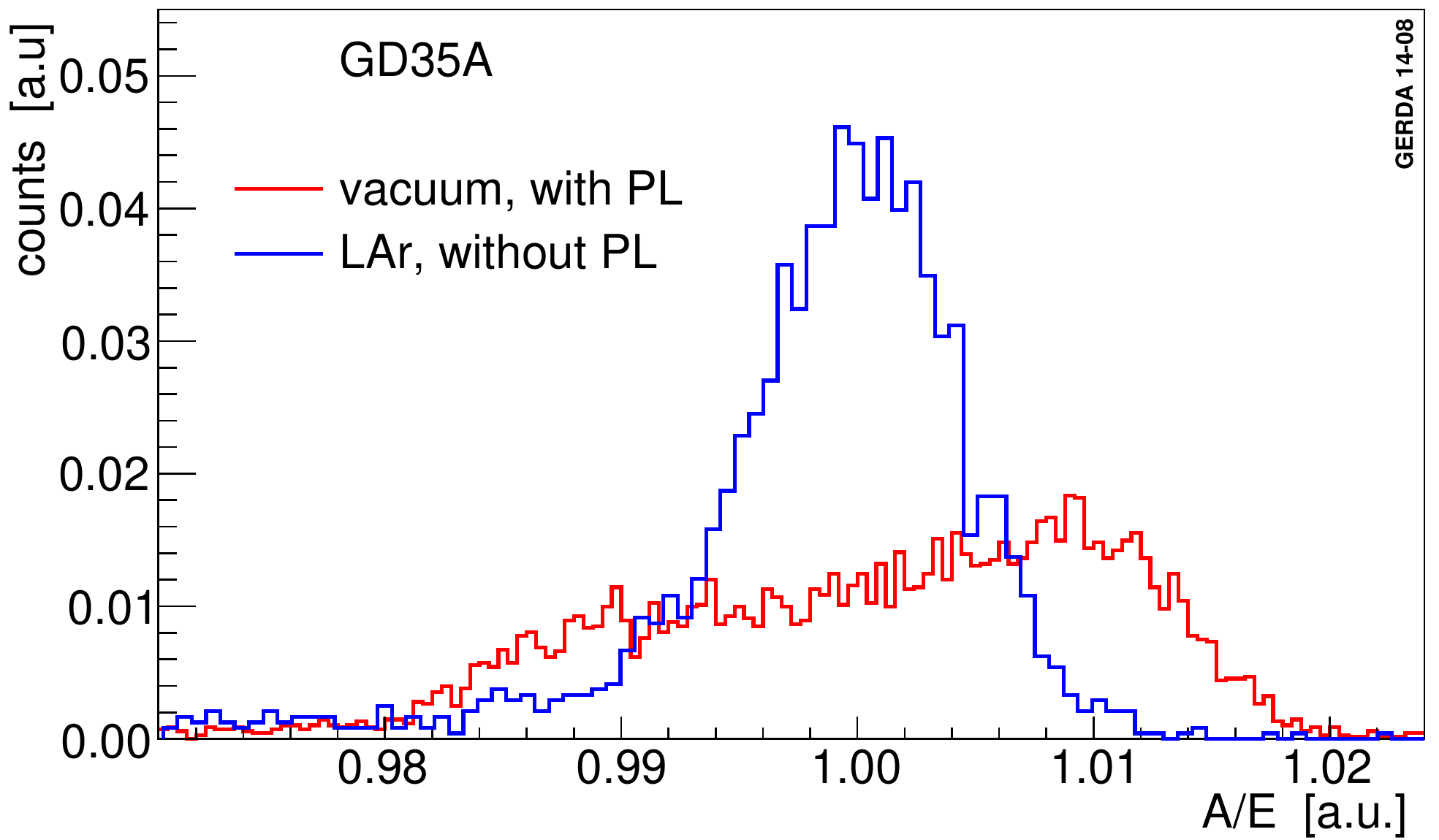}
  \caption{  \label{fig:GD35A_w-wo-PL}
         $A/E$ distributions of DEP events measured with detector GD35A under
         different conditions. Left: With passivation layer in vacuum. Right:
         Without passivation layer in LAr.
}
\end{figure}

The widths $b_{A/E}$ of the $A/E$ distributions of the two detectors GD32A and
GD35A improved from 1.3 and 2.4\% to 1.1 and 1.0\%, respectively. The residual
$A/E$ resolution was dominated by the noise of the GDL test set-up, and was
similar to past measurements with prototype BEGe detectors. The strong
non-Gaussian features observed in the HADES vacuum tests disappeared. As an
example, both $A/E$ distributions for detector GD35A are depicted in
Fig.~\ref{fig:GD35A_w-wo-PL}. The $A/E$ distribution measured in GDL is
consistent with a homogeneous pulse-shape response of the detector -- a
Gaussian SSE peak with a slight low $A/E$ tail due to bremsstrahlung.

The PSD survival efficiencies of $\gamma$-peaks and Comp\-ton-continua in
$^{228}$Th source spectra were determined for detector GD32A: The SEP, FEP and
Compton events at \qbb\ survive at 8.2, 11.5 and 38\%, respectively. It should
also be noted that for contaminants such as events from the detector n+
surface an improvement in $A/E$ resolution would have a greater effect on
reducing the survival efficiency than for MSE dominated background
contaminants such as $^{228}$Th.


\section{Conclusions}
\label{sec:conclusions}

This paper presents a detailed review of the production chain of the new
\gerda\ Phase~II broad energy germanium detectors.  It discusses the
performance of the first seven delivered detectors which were operated in
vacuum and in liquid argon.

The detector production included an efficient isotopic enrichment of $^{76}$Ge
from $\sim$8\% in natural germanium to $\sim$88\%, a successful purification
and crystal growth. 30 crystal slices were obtained and converted into
operational diodes. Only one diode did not fulfill all required performance
specifications by showing charge collection deficiencies in the AV. In total,
the 30 detectors have a mass of 20.0\,kg which corresponds to a mass yield of
53\%. The obtained impurity concentrations allowed for an operation of the
detectors at relatively low voltages of (3-4)\,kV. During all production
steps, the activation of the enriched germanium by cosmic radiation was kept
at a minimum in order to meet the background specifications of
\gerda\ Phase~II. This was pursued by storing the germanium in underground
sites near to the manufacturers and by using a shielded container for
transportation. According to the tracked exposure histories and the predicted
production rates of cosmogenic isotopes, the production of detectors
underground -- potentially needed in a future upgrade of the experiment for
further background reduction~\cite{bib:Hult2008} -- was not necessary at this
stage. Finally, more than 25\% of the original germanium material was
recovered for future crystal growth.

In order to characterize the detectors, an underground facility with the
capability of screening two detectors per week was installed. For the tests
the detectors were mounted in standard vacuum cryostats. As demonstrated by
means of the first seven detectors, they all turned out to have excellent
energy re\-solutions of 0.13\% in terms of full widths at half maximum of the
1333\,keV $^{60}$Co $\gamma$-line. This is an improvement of $\sim$30\%
compared to the \gerda\ detectors based on a semi-coaxial design. The active
volume fraction was measured with a precision of a few percent. The pulse
shape discrimination power was quantified. A previously unknown anomaly
originating probably from surface charges deposited in the groove around the
read-out electrode was identified. As a consequence, the PSD performance
notably deteriorated for some detectors. Nonetheless, possibilities for
improvement were found -- either by chemical and thermal treatment of the
passivation layer, or by removing the passivation entirely and handling the
detector under clean nitrogen atmosphere.

Five out of the seven enriched BEGe detectors were deployed in the LAr
cryostat of \gerda\ during Phase~I of the experiment. It was the first time
that this detector type was used in a \onbb\ decay experiment. All but one
were operated over almost 320\,d. The leakage current and the energy scale
were stable, while the energy resolution was $\sim$30\% worse compared to
their operation in vacuum. This was expected due to an increased distance of
the first signal amplifying stages from the detectors. The pulse shape
behavior experienced drifts in time which, however, could be corrected
offline. A similar, but dynamic mechanism as observed in vacuum could be the
reason for the effect. An appropriate modification of the passivation layer of
the detectors is expected to keep leakage currents low and to cure the
observed pulse shape degradations in LAr for \gerda\ Phase~II.

Beta-decays of the $^{42}$Ar daughter nuclide $^{42}$K, and gamma-rays induced
by $^{214}$Bi and $^{228}$Th decays were found to be the major contaminants
affecting the background region at \qbb\ for the BEGe detectors in \gerda. The
contribution from surface $\alpha$-events was of secondary order in the range
of 5\%. Moreover, the spectral fit of \gerda\ Phase~I data confirmed agreement
between the predicted and the fit-constrained contribution from cosmogenic
isotope decays. Signals induced by most types of backgrounds can be
efficiently removed by PSD cuts.  Further background rejection will be
possible by combining PSD with a veto by anti-coincidences between detectors
or by scintillation light in LAr induced by contaminants.  In addition, a
transparent nylon cylinder around the detectors to stop mechanically the
attracted $^{42}$K ions is in preparation. \gerda\ Phase~II is expected to
reduce its background by one order of magnitude compared to Phase~I. The
resulting background index of 10${^{-3}}$\,\ctsper\ in the region of interest,
the new additional detector mass and the improved energy resolution will allow
for the exploration of half life values above 10$^{26}$\,yr for the
\onbb\ decay of $^{76}$Ge after a few years of data collection.

\section*{Acknowledgments}
The authors are grateful to JSC PA Electrochemical Plant in Zelenogorsk, PPM
Pure Metals GmbH in Langelsheim, Canberra Industries Inc. in Oak Ridge, and
Canberra Semiconductor N.V. in Olen for the prolific cooperation and enduring
assistance.

 The \gerda\ experiment is supported financially by
   the German Federal Ministry for Education and Research (BMBF),
   the German Research Foundation (DFG) via the Excellence Cluster Universe,
   the Italian Istituto Nazionale di Fisica Nucleare (INFN),
   the Max Planck Society (MPG),
   the Polish National Science Centre (NCN),
   the Foundation for Polish Science (MPD programme),
   the Russian Foundation for Basic Research, and
   the Swiss National Science Foundation (SNF).
 The institutions acknowledge also internal financial support.

The \gerda\ collaboration thanks the \majorana\ collaboration for sharing the
underground site at Oak Ridge. Furthermore, we express our gratitude to the
team of EIG EURIDICE on the premises of the Belgian Nuclear Research Center
SCK$\bullet$CEN, Mol. We thank the directors and the staff of the LNGS
for their continuous strong support of the \gerda\ experiment.


\begin{thebibliography}{99}
\bibitem{bib:GER13} K.H. Ackermann et al.,
                 Eur. Phys. J. C {\bf 73} (2013) 2330  
\bibitem{bib:GER13-0nbb} M. Agostini et al.,
                   Phys. Rev. Lett. {\bf 111} (2013) 122503
\bibitem{bib:GER13-bckg} M. Agostini et al.,
               Eur. Phys. J. C {\bf 74} (2014) 2764 
\bibitem{bib:GER13-psd} M. Agostini et al.,
                Eur. Phys. J. C  {\bf 73} (2013) 2583
\bibitem{bib:qvalue} B. J. Mount, M. Redshaw, and E. G. Myers,
                  Phys. Rev. C    {\bf81} (2010) 032501
\bibitem{bib:HdM01} H.V. Klapdor-Kleingrothaus et al.,
                  Eur. Phys. J. A {\bf 12} (2001) 147-154 
\bibitem{bib:IGEX02} C.E. Aalseth et al.,
                  Phys. Rev. D {\bf 65} (2002) 092007
\bibitem{bib:bege_pulse_shape} D. Budj{\'a}{\v{s}} et al.,
             J. Instrum. {\bf 4} (2009) P10007  
\bibitem{bib:tipp}  M. Agostini et al.,
  procs. of TIPP in print, TIPP 2014: Internat. Conference on Technology and
  Instrumentation in Particle Physics, 2-6 June 2014, Amsterdam


\bibitem{bib:enrGe_k0naaa} P. Vermaercke et al.,
         Nucl. Instr. and Meth. A622 (2010) 433
\bibitem{bib:natGe-composition} J.R. De Laeter et al.,
                        Pure Appl. Chem. {\bf 75} (2003) 683
\bibitem{bib:deplBEGe13} D. Budj{\'a}{\v{s}} et al.,
                J. Instrum.  {\bf 8} (2013) P104018 

\bibitem{bib:ecp_ru} Joint Stock Company "Production Association
  Electrochemical Plant", Pervaya Promyshlennaya 1, 663690 Zelenogorsk,
  Russia; {\texttt{http://www.ecp.ru/index\_en.shtml}}
\bibitem{bib:enrGe-isotopic-comp} A. N. Shubin et al.,
         Atomic Energy {\bf 101} (2006) 588

\bibitem{bib:certCenterIMTHPM} Institute of Microelectronics Technology and
  High Purity Materials RAS, Academician Ossipyan Str 6, 142432 Chernogolovka,
  Russia;\\ {\texttt{http://www.ipmt-hpm.ac.ru/index.en.html}}
\bibitem{bib:giredmet} State Scientific-Research and Design Institute of
  Rare-Metal Industry "Giredmet" JSC, B. Tolmachevsky lane 5-1, 119017 Moscow,
  Russia;\\  {\texttt{http://giredmet.ru/en/instituteen/}}
\bibitem{bib:langelsheim} PPM Pure Metals GmbH, Am Bahnhof 1, 38685
  Langelsheim, Germany; {\texttt{http://www.ppmpuremetals.de/}} 
\bibitem{bib:canberra_usa} Canberra Industries Inc., 107 Union Valley Rd, Oak
  Ridge, TN, USA; {\texttt{http://www.canberra.com/}}
\bibitem{bib:pureGeXtal} E.E. Haller et al.,
                         Advances in Physics {\bf 30} (1981) 93 
\bibitem{bib:canberra_olen} Canberra Semiconductor N.V., Lammerdries 25, 2250
  Olen, Belgium ; {\texttt{http://www.canberra.com/}}


\bibitem{hayakawa1969} S. Hayakawa, Cosmic Ray Physics, John Wiley \& Sons,
                 Inc., NY (1969), ISBN-10: 0471363200, ISBN-13: 9780471363200
\bibitem{Ziegler1996} J.F. Ziegler, IBM J. Res. Dev. {\bf 40} (1996) 1
\bibitem{cebrian2010} S. Cebri{\'a}n et al.,
         Astropart. Phys. {\bf 33} (2010) 316 {\it and references therein}
\bibitem{barabanov2006} I. Barabanov et al.,
         Nucl. Instrum. Meth. B \textbf{251} (2006) 115
\bibitem{thesis_matteo} M. Agostini, PhD thesis,
           Technische Universit\"at  M\"unchen (2013)\\
{\texttt{http://www.mpi-hd.mpg.de/gerda/public/pub-phd.html}}

\bibitem{cosmic_rays_atm} P. Goldhagen et al.,
        Nucl. Instrum. Meth. A \textbf{476} (2002) 42 
\bibitem{diplom_aaron} A. Michel, diploma thesis, Technische Universit\"at
  M\"unchen (2012)


\bibitem{bib:shockley-ramo-theorem} Z. He,
          Nucl. Instrum. Meth. A {\bf 463} (2001) 250
\bibitem{bib:bege_pulse_simulation} M. Agostini et al.,
             J. Instrum. {\bf 6} (2011) P03005	
\bibitem{majorana2013} E. Aguayo et al.,
          Nucl. Instrum. Meth. A {\bf 701} (2013) 176
\bibitem{bib:pst_7500_SL}
  {\texttt{http://www.canberra.com/products/detectors/\\cryostats-coolers.asp}}
\bibitem{bib:gelatio1} M. Agostini et al., J. Instrum. {\bf 6} (2011) P08013
\bibitem{bib:gelatio2} M. Agostini et al.,
                                 J. Phys.: Conf. Ser. {\bf 368} (2012) 012047
\bibitem{bib:Heroica} E. Andreotti et al., J. Instrum. {\bf 8} (2013) P06012 

\bibitem{bib:Marik-thesis} M. Barnab{\'e}-Heider, PhD thesis,
              University of Heidelberg (2009)\\
{\texttt{http://www.mpi-hd.mpg.de/gerda/public/pub-phd.html}}
 \bibitem{bib:bubble-effect} M. Agostini et al.,
               J. Instrum. {\bf 6} (2011) P04005 
\bibitem{bib:pinch-off-effect} N. Abgrall et al.,
                Adv. High Energy Phys. {\bf 2014} (2014) 365432

\bibitem{bib:corrado} D. Budj{\'a}{\v{s}} et al.,
                 Appl. Rad. Isot. {\bf 67} (2009) 706
\bibitem{bib:MAGE} M. Boswell et al., IEEE Trans. Nucl. Sci. 58 (2011) 1212 
\bibitem{bib:GEANT4_1} S. Agostinelli et al.,
                  Nucl. Instrum. Meth. A {\bf 506} (2003) 250
\bibitem{bib:GEANT4_2} J. Allison et al.,
                  IEEE Trans. Nucl. Sci. {\bf 53} (2006) 270
\bibitem{bib:Geant4sys} G.A.P. Cirrone et al.,
                   Nucl. Instrum. Meth. A {\bf 618} (2010) 315


\bibitem{bib:NIST} NIST Photon Cross Section Database:\\
          {\texttt{http://www.nist.gov/pml/data/xcom/index.cfm}}
\bibitem{bib:Xtal-structure} I. Abt et al.,
                   Eur. Phys. J. C {\bf 68} (2010) 609


\bibitem{bib:BEGe-LAr_2010} M. Barnab{\`e}-Heider et al.,
                  J. Instrum. {\bf 5} (2010) P10007
\bibitem{bib:Marik2008} M. Barnab{\`e}-Heider et al.,
                   IEEE Conf. Nucl. Sci. Sym., October 2008, Dresden
\bibitem{bib:NISTbeta}  NIST Stopping-Power and Range Tables for Electrons,
                  Protons, and Helium Ions:\\
          {\texttt{http://www.nist.gov/pml/data/star/}}
\bibitem{bib:2012lazzaro} A. Lazzaro, Master thesis, University Milano Bicocca
                   (2012)


\bibitem{bib:Hult2008} M. Hult et al.,
           JRC Scien. Tech. Rep. (2008) EUR 23237 EN, ISBN: 978-92-79-08276-4 
\end{thebibliography}
\end{document}